\documentclass[aps,prb,twocolumn,groupedaddress,showpacs]{revtex4-1}  

\usepackage{graphicx}
\usepackage{bm}
\usepackage{color}
\usepackage{amsmath}
\usepackage{ifsym}
\usepackage{amssymb}

\newcommand{\degC}{$^{\circ}$C}
\newcommand{\degrees}{$^{\circ}$}
\newcommand{\FeCoSi}{Fe$_{1-x}$Co$_{x}$Si}

\begin{document}

\title{Scattering mechanisms in textured FeGe thin films:\\ magnetoresistance and the anomalous Hall effect}

\author{N.~A.~Porter}\email[email:~]{n.a.porter@leeds.ac.uk}
\affiliation{School of Physics \&\ Astronomy, University of Leeds, Leeds, LS2 9JT, UK}

\author{J.~C.~Gartside}
\altaffiliation[Present address: ]{Blackett Laboratory, Imperial College London, South Kensington Campus, London SW7 2AZ, UK}
\affiliation{School of Physics \&\ Astronomy, University of Leeds, Leeds, LS2 9JT, UK}

\author{C.~H.~Marrows}\email[email:~]{c.h.marrows@leeds.ac.uk}
\affiliation{School of Physics \&\ Astronomy, University of Leeds, Leeds, LS2 9JT, UK}



\begin{abstract}
  A textured thin film of FeGe was grown by magnetron sputtering with a helimagnetic ordering temperature of $T_\mathrm{N} = 276 \pm 2$~K. From 5~K to room temperature a variety of scattering processes contribute towards the overall longitudinal and Hall resistivities. These were studied by combining magnetometry and  magnetotransport measurements. The high-field magnetoresistance (MR) displays three clear temperature regimes: Lorentz force MR dominates at low temperatures, above $T \approx 80$~K scattering from spin-waves predominates, whilst finally for $T \gtrsim 200$~K scattering from fluctuating local moments describes the MR. At low fields, where the magnetisation is no longer technically saturated, we find a scaling of magnetoresistance with the square of the magnetisation, indicating that the MR due to the unwinding of spins in the conical phase arises from a similar mechanism to that in magnetic domain walls. This MR is only visible up to a temperature of about 200~K. No features can be found in the temperature or field dependence of the longitudinal resistivity that belie the presence of the underlying magnetic phase transition at $T_\mathrm{N}$: the marked changes in behavior are at much lower temperatures. The anomalous Hall effect has a dramatic temperature dependence in which the anomalous Hall resistivity scales quadratically with the longitudinal resistivity: comparison with anomalous Hall scaling theory shows that our system is in the intrinsic ``moderately dirty'' regime. Lastly, we find evidence of a topological Hall effect of size $\sim 100~\mu \Omega$cm.
\end{abstract}

\date{\today}
\maketitle

\section{introduction}

In recent years there has been a resurgence of interest in B20-ordered transition metal monosilicides\cite{Grigoriev2006,Muhlbauer2009, Karhu2010,Wilson2012,Li2013,Yu2010, Grigoriev2007,Porter2012} and monogermanides.\cite{Yu2011,Wilhelm2011,Huang2012,Moskvin2013, Shibata2013,Kanazawa2011,Kanazawa2012} The magnetic ground state of these alloys is helimagnetic by virtue of the hierarchy of energy terms which determine the magnetic order. What distinguishes these alloys from conventionally ordered ferromagnets is that the B20 unit cell lacks inversion symmetry. This results in a non-vanishing Dzyaloshinskii-Moriya interaction (DMI) which tends to favour orthogonal spin configurations. This is typically weaker than the ferromagnetic (FM) Heisenberg exchange which favours parallel spin alignment, and a compromise is reached causing neighbouring spins to cant with respect to one another leading to non-trivial chiral spin textures. In FeGe the ground state is helimagnetic with a pitch, $\lambda$, (determined by the ratio of the FM and DMI energy terms, $\lambda \propto A / D$) of $\sim$70~nm.\cite{Ericsson1981,Lebech1989,Yu2011,Shibata2013} When a magnetic field is applied along the propagation direction of the helix it is distorted into a conical state which saturates at high fields into the FM uniformly magnetized state. Neutron scattering\cite{Muhlbauer2009} provided clues as to a further complexity in the phase diagram leading to the direct observation\cite{Yu2010} of a complex spin modulation known as a skyrmion crystal: a hexagonally close-packed arrangement of topologically protected knot-like spin textures with potential applications in spintronics.\cite{Kiselev2011,Schulz2012,IwasakiNANO2013,Fert2013}

Much of the research on the B20 alloys has been on bulk single crystal material but there has been recent progress growing films by molecular beam epitaxy on Si~(111) substrates.\cite{Karhu2011,Porter2012,Sinha2014,Li2013} Epilayers of FeGe have recently been produced by Huang and Chien by the relatively faster and more cost-effective method of magnetron sputtering,\cite{Huang2012} a growth technique previously used to grow polycrystalline B20 \FeCoSi\cite{Morley2011}. Here we report the growth of a textured film of FeGe also using dc magnetron sputtering at elevated temperatures following a similar method to that described in Ref. \onlinecite{Huang2012}. In this report we focus on the magnetometry and magnetotransport of an 82~nm thick sputtered film, and have determined the scattering mechanisms responsible for the observed temperature dependence of the magnetoresistance (MR) and anomalous Hall effect (AHE).

Despite decades of research the AHE has many aspects that are still open questions.\cite{Nagaosa2010} The origins of the effect are all in some manner based upon spin-orbit coupling, but the individual mechanisms contributing to the overall effect can be difficult to deconvolute from one another. It is now generally accepted that skew scattering, side-jump scattering, and intrinsic anomalous contributions to the transverse resistivity $\rho_{xy}$ sum to account for the AHE, with each mechanism having a power law scaling with respect to the longitudinal resistivity $\rho_{xx}$.\cite{Nagaosa2010} Accurately determining which mechanism is most influential involves determining which scaling holds experimentally. This can be achieved by production of multiple samples either by doping,\cite{Manyala2004} tailoring thicknesses to alter the longitudinal resistivity,\cite{Tian2009,Ye2012} or simply from individual samples as was the case for early experiments on Fe and Ni.\cite{Karplus1954} In FeGe we measure here a single film which has a strong quadratic dependence of the AHE on the longitudinal resistivity over a wide temperature range.

In addition, we report a large MR in the saturated FM state that displays three clear temperature regimes in which different scattering mechanisms must predominate. An unexpected experimental fact is that none of the boundaries of these regimes coincide with the magnetic phase transition --- where there are, in fact, no discernible features in $\rho_{xx}(H,T)$ --- but occur at much lower temperatures.  By fitting the the high-field MR at all measured temperatures we are able to extract the MR due to scattering from the conical magnetic texture occurring below magnetic saturation, which arises from a similar giant magnetoresistance-type mechanism to that in magnetic domain walls. Finally, we observe the topological Hall effect in our FeGe film, reproducing the important observation of Huang and Chien,\cite{Huang2012} and confirming the quality of our thin film monogermanide materials.

\section{Sample growth and characterisation}

Textured FeGe films were grown by dc magnetron sputtering on substrates held at elevated temperatures with a base pressure of $1.3\times10^{-7}$~Torr. Pieces of Si (111) wafer (room temperature resistivity of 2-3 k$\Omega$cm) were annealed at 550~\degC\ for 5~hours then exposed to an {\it in-situ} Ar ion mill for 10~s to remove the native oxide. Approximately one atomic monolayer of pure Fe was then deposited to generate an FeSi seed layer by solid-phase epitaxy. The FeGe film was then co-sputtered from pure targets with a net rate of 1.2~\AA/s in an Ar:H$_2$ (4 \% H$_2$) working gas at 3~mTorr onto a substrate held at 470~\degC. The film thickness was determined by low angle X-ray reflectometry to be $82\pm2$~nm.

Textured FeGe (111) grows on Si (111) by virtue of a 30\degrees\ in-plane rotation of the interface FeGe with respect to the substrate.\cite{Chevrier1992} This has been demonstrated for other B20 alloys, such as MnSi\cite{Karhu2010,Karhu2011} and
\FeCoSi\cite{Porter2012}, with a biaxial strain of -3.0 and -5.6~\% respectively, distorting the cubic B20 phase into a rhombohedral form. By comparison, FeGe has only a -0.5~\% lattice mismatch with the substrate and as such the influence of strain should be much reduced.

\begin{figure}[tb]
  \includegraphics[width=8cm]{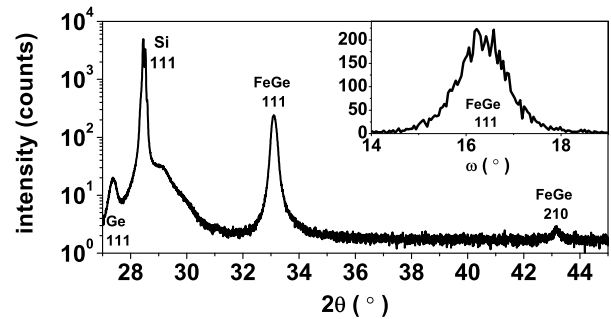}
  \caption{XRD of the 82~nm film has narrow Bragg peaks corresponding to the B20 phase of FeGe in the (111) texture and smaller impurity peaks arising from Ge and misoriented FeGe grains. A rocking curve taken at $2\theta = 33.1$\degrees\, through the FeGe (111) peak, is shown as an inset. \label{fig:XRD}}
\end{figure}

High angle X-ray diffraction (XRD) was used to determine the film texture, with the data displayed in Fig. \ref{fig:XRD}. Aside from the (111) reflection from the Si substrate, the most prominent Bragg reflection is from (111)-oriented FeGe, relating to an interplanar spacing of $d_\mathrm{111} = 2.704 \pm 0.001$ \AA. On the basis of the small strain relationship and assuming a cubic B20 unit cell with internal angles of 90\degrees\ the lattice parameter was hence calculated to be $a = 4.683 \pm 0.002$ \AA, approximately 0.1 \% larger than the lattice parameter reported for bulk of 4.679 \AA.\cite{Pedrazzini2007}

A rocking curve for the FeGe (111) reflection is shown in the inset of Fig. \ref{fig:XRD}, indicating a mosaic spread of $\sim 1$\degrees\, indicating a high degree of texture but not fully epitaxial growth. Although FeGe oriented with the (111) direction normal to the surface is the most prominent phase there was also a weak (210) reflection and some Ge (111) texture observed. The former is likely to arise in magnetron sputtering, which is a more energetic growth process than MBE, resulting in a small proportion of grains with different textures. The latter phase segregation is likely to be a result of slightly off stoichiometric growth. We will nevertheless see that the influence of this impurity phase on the magnetotransport properties of the textured film is minimal, and that FeGe plays the dominant role.

\section{Magnetometry}

\begin{figure}[tb]
  \includegraphics[width=8cm]{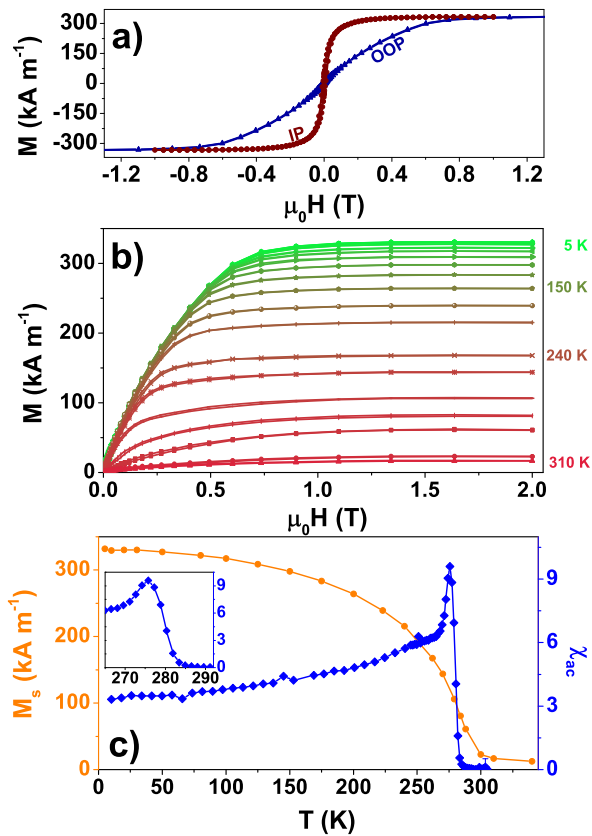}
  \caption{(Color online) Magnetic properties of the 82~nm FeGe film. (a) Anisotropy in the magnetization for in plane (IP) and out of plane (OOP) applied magnetic fields at 5~K. (b) Temperature evolution of the magnetization, measured in an OOP field, plotted from high (red) to low (green) temperatures with only a few labeled for clarity. (c) Saturation magnetization extracted  from the hysteresis loops (left ordinate axis) and magnetic susceptibility measured in an IP field, $\chi_\mathrm{ac}$, (right ordinate axis) in a dc magnetic field of 2~mT with an AC excitation of 1~mT at 23~Hz as a function of temperature. An enlarged view of $\chi_\mathrm{ac}$ close to $T_\mathrm{N}$ is shown as an inset.\label{fig:mag}}
\end{figure}

The magnetic properties of the film were characterized by superconducting interference device vibrating sample magnetometry (SQUID-VSM) in the temperature range 5-340~K, spanning the anticipated temperature for helimagnetic ordering reported for bulk of $T_{N} = 278.2$~K.\cite{Wilhelm2011} At 5~K, magnetic hysteresis loops were measured with the applied magnetic field, $H$, applied parallel (IP) or normal (OOP) to the film plane as shown in Fig. \ref{fig:mag}(a). The magnetization of $330 \pm 10$~kA/m corresponds to a moment of $m_\mathrm{s} = 0.924 \pm 0.003$~$\mu_\mathrm{B}$ per Fe atom, a figure that is slightly less than the bulk expectation of 1~$\mu_\mathrm{B}$ per Fe atom,\cite{Yamada2003, Pedrazzini2007,Huang2012} but nevertheless indicative that FeGe is the predominant alloy in the film.

Utilising the model of Karhu {\it et al.}\cite{Karhu2012} for films with conical modulations of the local magnetization, and assuming a helical wavelength of 70 nm\cite{Yu2011,Huang2012} an easy-plane magnetocrystalline anisotropy was calculated, with anisotropy constant $K_\mathrm{u} = 14.3$~kJ/m$^3$. For bulk FeGe the propagation direction, {\bf \^{n}}, of the helix is determined by a weak anisotropic exchange defined by the crystal axes,\cite{Lebech1989, Wilhelm2012} which favour propagation along the $\langle 111 \rangle$ directions at low temperatures.\cite{Uchida2008} This anisotropy can be easily overcome by a weak applied magnetic field which reorients the helix.\cite{Lebech1989} For our films both magnetocrystalline and shape anisotropy (for local spin directions) overshadow this weak crystal anisotropy (for the helix propagation direction), encouraging spins to lie within the plane of the film: this defines the propagation direction of the helix as being normal to the film plane. One would thus anticipate an in-plane field to unwind the helix with the spin remaining in the plane of the sample, giving rise to helicoidal states.\cite{Wilson2013} In an out-of-plane field the helix is expected to deform continuously from a helix through a conical state, finally reaching a uniformly magnetized (saturated) phase above a critical field, $H_\mathrm{c}$.\cite{Karhu2012}

The temperature evolution of the out-of-plane magnetization is shown in Fig. \ref{fig:mag}(b). The magnetization is weakly hysteretic, with a maximum coercive field of 10~mT at 5~K and less than 1~mT at 240~K, but otherwise the magnetization follows a nearly reversible sweep of the cone angle which decreases at higher fields approaching saturation. The saturation field $\mu_{0}H_\mathrm{c}$, where $\mu_{0}$ is the permeability of free space, is weakly temperature dependent and reduced from $\sim1$~T at 5~K to $\sim 750$~mT just below $T_\mathrm{N}$. The saturation magnetization $M_\mathrm{s}$ extrapolated from these data is plotted in Fig. \ref{fig:mag}(c). In order to accurately determine the ordering temperature $T_\mathrm{N}$, an IP field orientation was used with a 1~mT alternating magnetic field at 23~Hz applied in conjunction with a static field of 2~mT to determine the ac magnetic susceptibility $\chi_\mathrm{ac} = dM / dH$ of the film. The ordering temperature, $T_\mathrm{N} = 276 \pm 2$~K was determined from divergence of the susceptibility which is detailed in the plot that forms inset of Fig. \ref{fig:mag}(c). This value falls within the range of reported values for bulk of 275-278.7~K.\cite{Ericsson1981,Lebech1989,Wilhelm2011}

At high temperatures, the susceptibility is found to obey a Curie-Weiss law, $\chi = C /(T - T_\mathrm{N})$. Fitting the data yields $T_\mathrm{N} = 283 \pm 1$~K, slightly higher than that determined above, and a Curie constant $C = 0.40 \pm 0.03$~K. This gives a moment of $m_\mathrm{c} = \sqrt{3k_\mathrm{B}C/\mu_0 n} = 2.8 \pm 0.1$~$\mu_\mathrm{B}$ per Fe atom in the paramagnetic phase, where $n$ is the number density of Fe atoms. This yields a ratio $m_\mathrm{c}/m_\mathrm{s} = 3.1 \pm 0.1$, implying a considerable degree of itinerancy in the moments according to the Rhodes-Wohlfarth picture.\cite{Rhodes1963}

\section{Longitudinal resistivity}\label{sec:Rxx}

\subsection{Temperature dependence}

\begin{figure}[tb]
  \includegraphics[width=6.5cm]{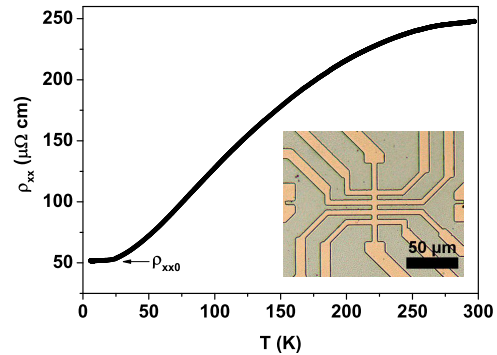}
  \caption{(Color online) Resistivity $\rho_{xx}$ of the patterned Hall bar device (image from an optical microscope inset), measured at zero field as a function of temperature. The residual resistivity $\rho_{xx0} \approx 50 \mu\Omega$cm is indicated by an arrow.\label{fig:rhoXXvT}}
\end{figure}

The thin film was then patterned using photolithography and Ar ion milling to define a Hall bar with a 5~$\mu$m width. A micrograph of the final device is shown as the inset of Fig. \ref{fig:rhoXXvT}. The longitudinal resistivity $\rho_{xx}$ was measured using a four wire method, biased with a dc current (used for all subsequent transport) of $\pm 100~\mu$A. The data depicted in Fig. \ref{fig:rhoXXvT} show the temperature dependence of the resistivity, which is similar in size and functional form to previous FeGe films.\cite{Huang2012} Although the behavior is metallic, in the sense that $d\rho_{xx} / dT$ is positive at all temperatures,  there are two features that depart from the usual properties of a magnetically ordered metal: the curve is not linear at high temperatures, and there is no cusp (or feature of any sort) at the magnetic ordering temperature $T_\mathrm{N}$.

\subsection{High-field magnetoresistance}

\begin{figure}[tb]
  \includegraphics[width=8cm]{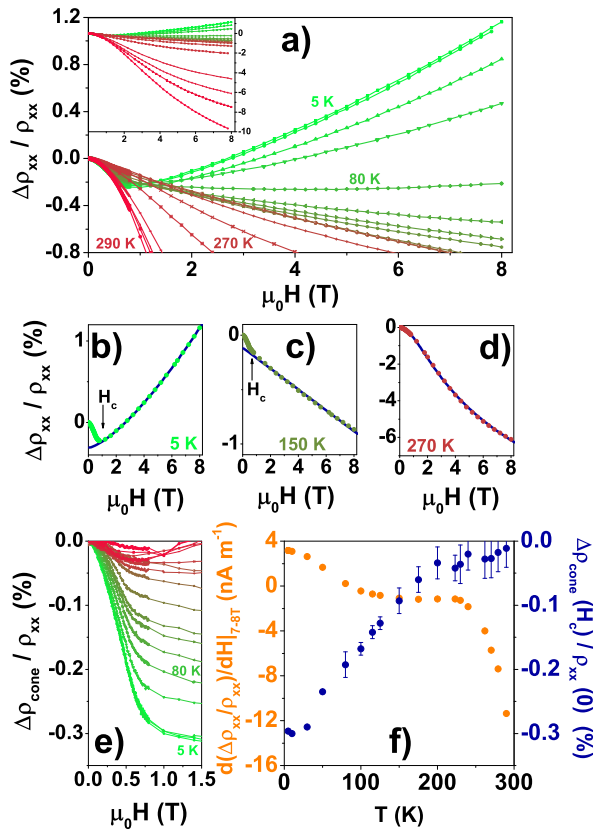}
  \caption{(Color online) Magnetoresistance of the Hall bar device at various temperatures measured with the magnetic field applied out-of-plane. (a) MR$(H)$ at various temperatures. In high fields (above $H_\mathrm{c}$) the MR varies from a positive contribution below 80~K to a large negative contribution near room temperature (see inset). (b), (c), and (d) show fits (solid lines) to the MR for $H > H_\mathrm{c}$ at 5, 150, and 270~K respectively, using models described in the main text. (e) The contribution to MR in the conical phase extracted the data shown in panel (a) by subtraction of the high field MR. (f) The gradient of the MR between 7 and 8~T is shown on the left  ordinate axis. The large negative MR persists even above the ordering temperature. The right ordinate axis illustrates the residual MR from the fits originating from the conical phase.\label{fig:rhoXXvH}}
\end{figure}

The MR was measured with the magnetic field applied perpendicular to the sample plane, and is shown in Fig. \ref{fig:rhoXXvH}(a). For temperatures below about 200~K, there is a low field contribution to the MR, $\rho_\mathrm{cone}$, which is the contribution to scattering arising from the conical magnetic state and saturates beyond $H_\mathrm{c}$. This has been shown for an in-plane field in MnSi where the helix is discontinuously unwound into a low resistance saturated state.\cite{Wilson2013} With the helical propagation vector of the helix defined by the anisotropy one would expect a continuous rotation of the high resistance spin helix through the conical state to the low resistance saturated state, leading to a smooth reduction in $\rho_{xx}$up to $H = H_\mathrm{c}$, exactly as is seen in our data. We will discuss this conical magnetoresistance in more detail in \S\ref{sec:conemr} below.

Beyond $H_\mathrm{c}$, the magnetisation is technically saturated and other high-field mechanisms will come into play. It is helpful to consider these scattering contributions in three temperature regimes. The $T$ dependence high field slope of the MR is displayed in Fig. \ref{fig:rhoXXvH}(f), exhibiting two marked changes in slope, which separate these three regimes. Below about 80~K the high-field MR is positive, with positive curvature. Above this temperature the MR switches over to be negative: initially it is close to linear, but then goes on to develops a pronounced positive curvature once again at temperatures above about 200~K. Evidently, different underlying mechanisms of high-field MR are predominant in these different regimes.

The MR at low temperatures and high fields ($H > H_\mathrm{c}$), defined as $\Delta\rho_{xx}/\rho_{xx} = (\rho_{xx}(H) - \rho_{xx}(0)) / \rho_{xx}(0)$, originates from the orbital motion of free carriers due to the Lorentz force, and depends upon the Larmor frequency $\omega_\mathrm{c}$ and the mean free scattering time $\tau$. This contribution to the longitudinal resistivity takes the form $\rho_{xx} \propto (\omega_\mathrm{c}\tau)^q = (\mu\mu_{0} H)^q$, where $\mu$ is the carrier mobility and the exponent $q = 2$ in standard theories that consider cyclotron motion of the electrons in the field.\cite{PippardBook1989} However, departures from $q=2$ are commonplace in the experimental literature, with $1<q<2$ having been observed in systems as diverse as spin-glasses,\cite{Yosida1957}, degenerately doped semiconductors,\cite{Khosla1970} and thin films of ferromagnetic metals.\cite{Raquet2002} As the temperature rises, the mean free scattering time shortens, weakening this orbital MR contribution. Ferromagnets are known to display a negative linear MR, arising from the suppression of electron-magnon scattering when a high field opens a gap in the magnon spectrum.\cite{Raquet2002,Marrows2004} This effect gets stronger as the temperature rises, since there are then more magnons to suppress. The observed MR at intermediate temperatures is as we would expect based on this picture when the film is in a single domain FM state (\textit{i.e.} $H > H_\mathrm{c}$). This linear MR is expected to be valid for $T < T_\mathrm{N}/2$ but for $T > T_\mathrm{N}/2$ this physical picture is no longer applicable as it underestimates spin disorder (just as conventional spin wave theory, which yields the low temperature $T^{3/2}$ Bloch law, fails as the critical temperature is approached). This spin disorder becomes significant near $T_\mathrm{N}$ where optical magnons and Stoner excitations\cite{Ishikawa1977} are thermally populated, providing short range spin fluctuations. The MR in this regime has yet to receive detailed theoretical attention. Above $T_\mathrm{N}$ only the local moments that contribute to the paramagnetic susceptibility persist, where we observe a negative, nonlinear MR. Extending the work of Yosida,\cite{Yosida1957} Khosla and Fischer have given a semi-empirical model of local moment scattering based on third order perturbation expansions of the $s$-$d$ exchange Hamiltonian, which predicts this form of the MR.\cite{Khosla1970} Although developed to describe scattering from In impurities in CdS, it has subsequently been widely applied to many other materials systems. Given that the essential ingredients of the model are simply a degenerate electron gas containing thermally fluctuating local moments, and so we can expect that it can also describe our data at high temperatures.

For $T \lesssim 80$ K, the orbital MR dominates at high fields and the longitudinal resistivity is expected to obey
\begin{equation}\label{eqn:MRlowT}
  \Delta\rho_{xx}(H) = \Delta\rho_\mathrm{cone}(H) + \rho_{xx}(0)(\mu \mu_{0}H)^q,
\end{equation}
where $\mu$ is the mobility For $80 \lesssim T \lesssim 200$ K, the linear, negative MR from electron-magnon scattering is the dominant high field contribution. In this regime the total MR is described to given by
\begin{equation}\label{eqn:MRmidT}
  \Delta\rho_{xx}(H) = \Delta\rho_\mathrm{cone}(H) - c_{1} \mu_{0}H.
\end{equation} For $T \gtrsim 200$ K spin-fluctuations become important and the semi-empirical formula of Khosla and Fischer\cite{Khosla1970} for the local moment scattering,
\begin{equation}\label{eqn:MRhighT}
  \Delta\rho_{xx}(H) = \Delta\rho_\mathrm{cone}(H) - b_{1}\ln\left(1 + (b_{2}\mu_{0}H)^2\right),
\end{equation}
best describes the high-field MR.

Representative fits of the MR for $H > H_\mathrm{c}$ in these three temperature regimes (using Eqs \ref{eqn:MRlowT}, \ref{eqn:MRmidT}, and \ref{eqn:MRhighT}) are shown in Figs \ref{fig:rhoXXvH}(b), \ref{fig:rhoXXvH}(c), and \ref{fig:rhoXXvH}(d) respectively. The parameters returned by these three fits were (at 5~K) $\mu = 0.0064$~m$^2/$Vs and $q = 1.41$, (at 150~K) $c_{1} = 0.18~\mu\Omega$cm$/$T, and (at 270~K) $b_{1} = 4.9~\mu\Omega$cm and $b_{2} = 0.56$~m$^2/$Vs. As in many materials, we do not obtain the canonical result of $q = 2$. Rather, we find here that $1 < q < 2$, as reported in ferromagnetic iron,\cite{Taylor1968,Raquet2002} for instance. Although the magnon scattering MR in metallic Fe, Co, and Ni was found to fall on a universal scaling curve for all three metals, the value of $c_1$ we obtain here is roughly an order of magnitude too large to scale in the same way. This points to a different form of magnetic excitation in FeGe, not surprising in view of the helimagnetic, rather than ferromagnetic, ground state of this material. The fitting parameters returned by the Khosla-Fischer expression are not straightforward to interpret, although we note that our value for $b_2$ is small compared to those reported in their paper for CdS:In, consistent with the trends they observe in carrier density and temperature.

As shown in the inset of Fig. \ref{fig:rhoXXvH}(a), the MR becomes dramatically stronger near the ordering temperature. This is summarised in Fig. \ref{fig:rhoXXvH}(f), where the field derivative of the MR between 7 and 8~T is plotted as a function of temperature. The changes in gradient of this curve clearly reveal the three temperature regimes described above. A dramatic enhancement begins at $\sim 230$~K and extends well beyond $T_\mathrm{N} = 276$~K. It is remarkable that, like the $\rho_{xx} (T)$ curve presented in Fig. \ref{fig:rhoXXvT}, there is no change in the MR that indicates that a critical point has been passed and a phase transition has taken place at $T_\mathrm{N}$. Indeed, it is very surprising that the Khosla-Fischer formula still fits the data very well for $T < T_\mathrm{N}$, where it is not obvious that the assumptions that underpin it still hold good. Whilst this may be fortuitous, it is also possible (given the lack any indication of a phase transition in the magnetotransport) that the scattering processes in FeGe not too far below $T_\mathrm{N}$ are quite similar to those in the paramagnetic regime.

Within the limits for our measurement equipment we found that the MR is still increasing for $T$ exceeding $T_\mathrm{N}$ by as much as $14 \pm 3$~K. It is likely that this increase in MR is due to the spin fluctuations that are expected near to critical points, where one expects low energy fluctuations with extended correlation lengths.\cite{Janoschek2013} Near to the ordering temperature these spin fluctuations may be helical in nature due to the presence of DMI, but tend to be FM in character further from $T_\mathrm{N}$, all the while reducing in intensity.\cite{Janoschek2013} Thus at higher temperatures the influence of spin fluctuations on the resistivity is diminished and we would anticipate the magnitude of this contribution to decrease again for $T > T_\mathrm{N}$, as is the case with helimagnet MnSi.\cite{Kadowaki1982,Sakakibara1982,Demishev2012} In FeGe, magnetic fluctuations have been measured well above the temperatures for establishing long-ranged magnetic order,\cite{Wilhelm2012} which can account for the observed MR far above $T_\mathrm{N}$ in our film.

\subsection{Low-field magnetoresistance}
\label{sec:conemr}

Below $H_\mathrm{c}$ the MR arises from the closing up of the moments in the field direction: the zero-field helical state is deformed into a conical state and the moments all align with the field direction beyond $H_\mathrm{c}$. Using the fits from Eqs \ref{eqn:MRlowT}, \ref{eqn:MRmidT}, and \ref{eqn:MRhighT} the contribution to the MR from these high field effects was subtracted from the data to yield the MR arising from the suppression of the conical state, $\Delta\rho_\mathrm{cone} / \rho_{xx}$: isotherms are plotted in Fig. \ref{fig:rhoXXvH}(e). The resistance is highest at zero field in the helical state and saturates above $H_\mathrm{c}$. The magnitude of this MR at saturation is plotted in Fig. \ref{fig:rhoXXvH}(f) as a function of temperature. The contribution to scattering from the conical phase is weakest near $T_\mathrm{N}$, but as the temperature is reduced scattering from the spin texture becomes a significant contribution to the film resistance. It is remarkable that this MR arising from the suppression of the conical state collapses at $\sim 200$~K, well below $T_\mathrm{N}$. This lends further support to the idea that the scattering processes in the regime 200~K~$\lesssim T < T_\mathrm{N}$ are akin to to those in the paramagnetic regime, but are distinct from those which predominate at lower temperatures.

The conical MR mechanism is very similar to that for domain wall MR,\cite{Marrows2005} where spin misalignments on neighbouring atomic sites lead to mixing of the spin channels. This can be treated quantum mechanically using a ``giant-magnetoresistance'' (GMR)-type Hamiltonian,\cite{Levy1997} where the domain wall is treated as half a turn of a spin helix. In this case a scaling of $\Delta \rho_\mathrm{cone} \sim - M^2$ is expected.

In a GMR-type model, the MR can be written as
\begin{equation}
\frac{\Delta \rho_\mathrm{cone}}{\rho_{xx}} = - \left( \frac{\Delta \rho_\mathrm{max}}{\rho_{xx}} \right) \times \frac{1}{2} \left[ 1 + \cos \psi \right] \label{eqn:coneMR}
\end{equation}
when normalized to the zero-field value of $\rho_{xx}$, where $\psi$ is the angle between spins in neighboring atomic planes and $\Delta \rho_\mathrm{max}$ is the change in resistivity when neighbouring spins go from fully antiparallel to parallel under the application of an applied field.

It is convenient to work in spherical co-ordinates. In the conical state the helical modulation vector lies in the field direction, and so all moments make an angle $\theta$ with the field axis (the cone angle), which drops from $\pi/2$ to zero as the field is increased. Thus, $\cos \theta = \left( M / M_\mathrm{s} \right)$.

Along the helical direction, there will be a difference in azimuthal angle $\Delta \phi = 2 \pi a/ \lambda$ between spins on adjacent lattice sites (where $a$ is the lattice constant and $\lambda$ is the period of the spin helix). Writing the azimuthal angle of one moment $\mathbf{m}_1$ as $-\Delta \phi /2$ and its neighbor $\mathbf{m}_2$ as $\Delta \phi /2$, we can express the scalar product of the two spins as

\begin{widetext}
\begin{equation}
\textbf{m}_1 \cdot \textbf{m}_2 =  m^2 \left[ \sin^2 \theta \cos^2 \left( \frac{\Delta \phi}{2}  \right)  - \sin^2 \theta \sin^2 \left( \frac{\Delta \phi}{2} \right) + \cos^2 \theta \right]. \label{eqn:conical}
\end{equation}
Inspection of Eq. \ref{eqn:conical} shows that the expression within square brackets is equal to $\cos\psi$. Defining $f = \cos^2 \left( \Delta \phi /2 \right) - \sin^2 \left( \Delta \phi /2 \right) = \cos \Delta \phi)$, we can see that $\cos \psi = f + (1-f) \cos^2 \theta$, where the quantity $f$ gives the reduction in MR amplitude due to longer period modulations with respect to the maximum that occurs case where neighboring spins lie antiparallel in the ground state, i.e. $\lambda = 2a$.

This result can be substituted into Eq. \ref{eqn:coneMR} to obtain the required scaling of MR with $-M^2$ in the only term that varies with magnetic field:
\begin{equation}
- \left( \frac{\Delta \rho_\mathrm{cone}(H)}{\rho_{xx}} \right) = \left( \frac{\Delta \rho_\mathrm{max}}{\rho_{xx}} \right) \times \frac{1}{2} \left[ 1 + \left\{ f + (1-f) \left( \frac{M(H)}{M_\mathrm{s}} \right)^2 \right\} \right]. \label{eqn:coneMR2}
\end{equation}
\end{widetext}

\begin{figure}[tb]
  \includegraphics[width=8cm]{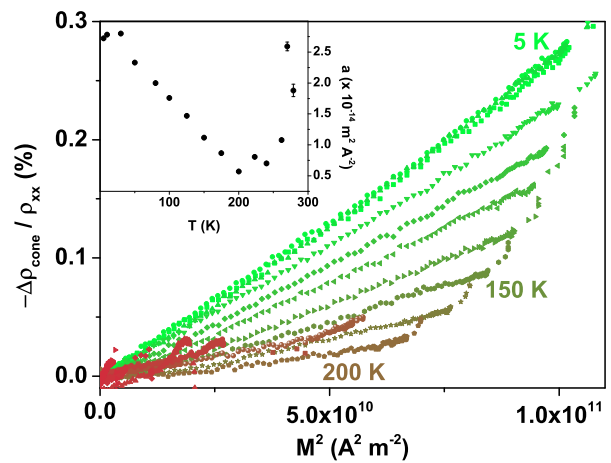}
  \caption{(Color online) Scaling of the magnetoresistance in the conical state. Below the critical field $H_\mathrm{c}$, above which the magnetisation is uniform, there is a clear scaling of the form $ - \left( \Delta \rho_\mathrm{cone}/ \rho_{xx} \right) = a M^2$, as predicted in a GMR-type model. The inset shows the temperature dependence of the scaling constant $a$.  \label{fig:scaling}}
\end{figure}

A plot of conical MR (the data for $-(\Delta \rho_\mathrm{cone}/\rho_{xx})$ shown in Fig. \ref{fig:rhoXXvH}(e)) against $M^2$ is shown in Fig. \ref{fig:scaling} for various temperatures, confirming that this simple model captures the relevant physics. In all cases up to $T \approx 200$~K the data show excellent linearity up to $H_\mathrm{c}$, beyond which, as expected, there is a sharp upturn in the data and the scaling breaks down as the conical state has now been saturated. The scaling constant $a = (-\Delta \rho_\mathrm{cone}/ \rho_{xx})/M^2$ drops linearly with temperature up to 200~K. At higher temperatures the very small conical MR means that the data are quite noisy, but a roughly linear behavior can still be discerned. From the present data set, it is not possible to decompose the variation in $a$ with temperature into that part arising from changes in the helical pitch (which will modify $f$ in Eq. \ref{fig:scaling}) and changes in the prefactor $(\Delta \rho_\mathrm{max}/\rho_{xx})$ (which arises from nonadiabatic spin-mistracking in the Levy-Zhang theory\cite{Levy1997}). A separate measurement of one or the other is required, although the latter can be expected to have a much stronger temperature dependence, since it depends on the spin-polarization of the current,\cite{Levy1997,Marrows2004} which drops steeply as the temperature rises.

\section{Anomalous Hall effect}\label{sec:AHE}

\begin{figure}[tb]
  \includegraphics[width=8cm]{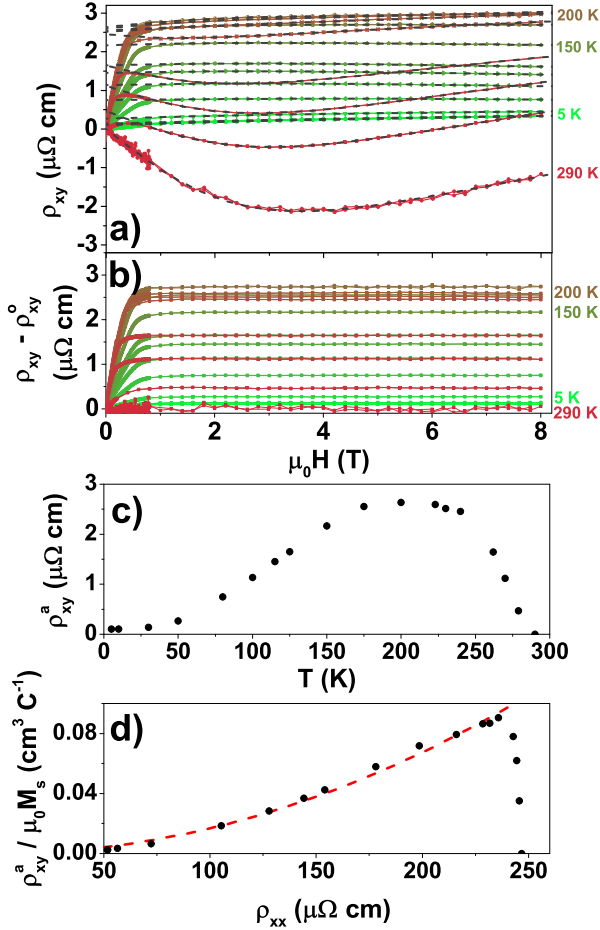}
  \caption{(Color online) The Hall effect. (a) The Hall resistivity $\rho_{xy}$ of the device is shown measured in the temperature range 5-290~K. The dashed lines are fits to the ordinary Hall effect using the two carrier model. (b) The Hall effect from (a) with ordinary Hall contributions subtracted leaving only anomalous and topological contributions. (c) The strong temperature dependence of the anomalous component of the Hall effect $\rho_{xy}^\mathrm{a}$
  peaks near 200~K. (d) Correlation of the Hall  resistivity scaled by the saturation magnetization with the longitudinal resistivity. The dashed line is a fit of the quadratic function given by Eq. \ref{eqn:AHE1} with $\alpha = 0$ to the data for $T < 230$~K.\label{fig:rhoXY}}
\end{figure}

The temperature dependence of the off-diagonal components of the resistivity tensor were measured in the Hall voltage, which was measured simultaneously with the MR. The Hall resistivity $\rho_{xy}$ is shown in Fig. \ref{fig:rhoXY}(a). In helimagnets it is expected to comprise three contributions:\cite{Nagaosa2012}
\begin{equation}\label{eqn:AllHall}
  \rho_{xy} = \rho_{xy}^\mathrm{o} + \rho_{xy}^\mathrm{a} + \rho_{xy}^\mathrm{t}.
\end{equation}
The first term, the ordinary Hall effect (OHE), $\rho_{xy}^\mathrm{o} = R_{0}\mu_{0}H$, is proportional to the applied field. The second term $\rho_{xy}^\mathrm{a} = R_\mathrm{s}M$ is the anomalous Hall effect (AHE) contribution arising from spin-orbit coupling, and is proportional to the magnetization $M$ through the anomalous Hall coefficient $R_\mathrm{s}$.\cite{Nagaosa2010} The third term $\rho_{xy}^\mathrm{t}$ is the topological Hall effect (THE), which arises from Berry phase effects when the spin textures in the film exhibit non-zero skyrmion winding number density.\cite{Bruno2004,Tatara2008} This has previously been detected in FeGe thin films,\cite{Huang2012} and will discussed in \S\ref{sec:THE} below.

At low temperatures the Hall resistivity looks as expected from this simple analysis: there is a steep rise at low fields due to the anomalous and any topological  contribution, and then a shallow slope when only the ordinary part remains. At 5~K, this ordinary Hall slope corresponds to an electron density of $(2.41 \pm 2) \times 10^{22}$~cm$^{-3}$. Nevertheless, above about 200~K there is a nonlinearity in the ordinary Hall effect shown in Fig. \ref{fig:rhoXY}(a), even above the saturation field. This is particularly evident at 290 K where, although the MR (see Fig. \ref{fig:rhoXXvH}(f)) is at its strongest, the influence of the anomalous Hall resistivity arising from the longitudinal resistivity is far too small to account for the observed nonlinearity. This nonlinear Hall signal can arise in the situation where there are two carrier types ($j = 1,2$) with Hall coefficients, $R_{j}$, and differing mobilities, $\mu_{j}$, such that each channel contributes $\rho_{j}^{-1} = \mu_{j}/R_{j}$ to the conductivity.\cite{HurdHallBook1972,Kim1999} If $f_{j} = \rho_{j}^{-1} / \rho_{xx}^{-1}$ is the fraction each carrier contributes to the longitudinal conductivity (such that $f_{1}+f_{2}=1$), then the net Hall coefficient is\cite{Kim1999}:
\begin{equation}\label{eqn:TwoCarrier}
R_{0} = \rho_{xx}\frac{A_{2} + B_{2}(\mu_{0}H)^2}{1 + B_{3}(\mu_{0}H)^2},
\end{equation}
where, $A_{2} = f_{1}\mu_{1} + f_{2}\mu_{2}$, $B_{2} = \mu_{1}\mu_{2}(f_{1}\mu_{2} + f_{2}\mu_{1})$ and $B_{3} = (f_{1}\mu_{2} + f_{2}\mu_{1})^2$.

Although solutions for the individual mobilities and carrier concentrations cannot be extracted from fits to the Hall resistivity alone, Eq. \ref{eqn:TwoCarrier} was used to fit the non-linear Hall resistivity above $H_\mathrm{c}$ provided $\rho_{xy}^\mathrm{o} = R_{0}\mu_{0}H$. The excellent fits to the data are shown in Fig. \ref{fig:rhoXY}(a). This technique to account for the OHE has been used recently to obtain the THE in Mn$_{1-x}$Fe$_{x}$Si films\cite{Yokouchi2014}. Of course the detailed band structure of FeGe is unknown at the present time, and this two-carrier model is likely to be a simplification. However, the transport being dominated by a small number of the many bands that cross the Fermi level is quite plausible.

These fits provide a sufficiently accurate empirical description of the OHE background that they were used to substract it from the raw data in Fig. \ref{fig:rhoXY}(a). Fig. \ref{fig:rhoXY}(b) depicts the remaining Hall signals after this subtraction. The temperature dependence of this data is summarised in Fig. \ref{fig:rhoXY}(c) at saturation for $H > H_\mathrm{c}$ where one would expect no topological contributions and thus $\rho_{xy}^\mathrm{a} = \rho_{xy} - \rho_{xy}^\mathrm{o}$. This highlights the strong temperature dependence of the AHE. From 5-200~K the AHE increases by over an order of magnitude, to a peak value of 2.6~$\mu \Omega$cm at about 200~K, but then drops dramatically as $T_\mathrm{N}$ is approached, where the magnetization is diminished ({\it c.f.} the behavior showed in Fig. \ref{fig:mag}(c)).

In the remainder of this section we determine the origin of this AHE, using the usual scaling methods. The AHE in general has three contributing terms:\cite{Nagaosa2010}
\begin{equation}\label{eqn:AHE1}
\rho_{xy}^\mathrm{a} = \left[ \alpha\rho_{xx} + \beta\rho_{xx}^2 + b\rho_{xx}^2
\right] \mu_{0}M(H).
\end{equation}
The terms with coefficients of $\alpha$, $\beta$, $b$ correspond to skew scattering, side-jump
scattering, and intrinsic ($k$-space Berry curvature) contributions, respectively. A similar temperature dependence of the AHE shown in Fig. \ref{fig:rhoXY}(c) was measured in Mn$_{5}$Ge$_{3}$,\cite{Zeng2006} which the authors attributed to the intrinsic AHE. In Fig. \ref{fig:rhoXY}(d) we plot the AHE scaled by the saturation magnetization, $\rho_{xy}^\mathrm{a} / (\mu_{0}M_\mathrm{s})$, and find that sufficiently far below the ordering temperature ($T < 230$~K) only $\rho_{xx}^2$ terms are required to describe changes in the AHE: a fit of Eq. \ref{eqn:AHE1} with $\alpha$ set to zero is shown by the dashed line in Fig. \ref{fig:rhoXY}(d). (Including a linear term in the fit barely changes the result, with the coefficient of that term returned by the fit being tiny.) Thus we find that the skew scattering contribution, which is typically only significant in nearly perfect crystals\cite{Onoda2008,Nagaosa2010}, is negligible. Nevertheless, the quadratic scaling that we observe is consistent with both the intrinsic (Berry phase) and side-jump scattering mechanisms. At low temperatures, the measured anomalous Hall and longitudinal resistivities imply that $\sigma_{xx} = \rho_{xx}/(\rho_{xx}^2 + \rho_{xy}^2) = 1.94 \times 10^{4}~(\Omega \mathrm{cm})^{-1}$, whilst $|\sigma_{xy}|= |\rho_{xy}|/(\rho_{xx}^2 + \rho_{xy}^2) = 3.83 \times 10^{1}~(\Omega \mathrm{cm})^{-1}$. This falls onto the scaling curve in the theory of Onoda, Sugimoto, and Nagaosa squarely within the `moderately dirty' regime,\cite{Onoda2008} where $\sigma_{xy}$ is independent of $\sigma_{xx}$: just the scaling we see as the temperature is varied. Onoda \textit{et al.} argue that the intrinsic mechanism dominates over side-jump scattering in that regime.

\section{Topological Hall effect}\label{sec:THE}

These results on the scattering leading to MR and the AHE are the main experimental results we report here. However, Eq. \ref{eqn:AllHall} alludes to the possibility of other contributions to the total Hall effect that cannot be accounted for by ordinary and anomalous contributions alone. This component $\rho_{xy}^\mathrm{t}$ is known as the topological Hall effect, previously studied in FeGe thin films by Huang and Chien.\cite{Huang2012} In order to elucidate the field dependence of the THE we calculate the sum of the ordinary and anomalous contributions (following the same standard procedure as Refs. \cite{Kanazawa2011,Huang2012,Li2013}) accounting for the hysteresis in $M(H)$ that will be reflected in the field dependence of the AHE (Eq. \ref{eqn:AHE1}), as well as including a term linear in field to account for the OHE:
\begin{equation}\label{eqn:HallEta}
    \begin{split}
     \eta(H) & =  \rho_{xy}^\mathrm{a} + \rho_{xy}^\mathrm{o}  \\
    & =  (\beta + b)\rho_{xx}^2\mu_{0}M(H) + \mu_{0}R_{0}H.
    \end{split}
\end{equation}

\begin{figure}[tb]
\includegraphics[width=8cm]{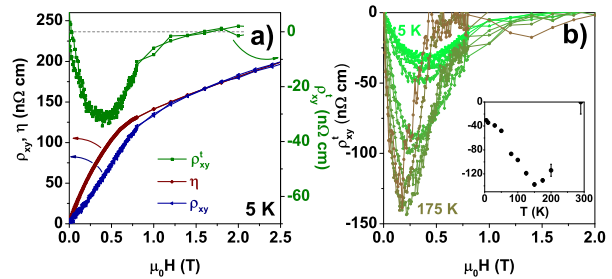}
  \caption{(Color online) Topological Hall effect (THE). (a) Topological contribution to the Hall   effect $\rho_{xy}^\mathrm{t}$ at $T = 5$~K. The discrepancy between the Hall resistivity $\rho_{xy}$ and the sum of the anomalous and ordinary contributions, $\eta$, is accounted for by THE, plotted on the  (right-hand ordinate axis). (b) Temperature evolution of the THE up to 175 K with the temperature dependence of the extremum shown inset.\label{fig:THE}}
\end{figure}

This was fitted to the total Hall effect for fields above $H_\mathrm{c}$ (where a uniformly magnetised state is expected and thus $\rho_{xy}^\mathrm{t}$ should vanish) and is plotted in Fig. \ref{fig:THE}(a) alongside the Hall resistivity data. There is a clear discrepancy between the two curves, which is attributed to the THE. This difference, $\rho_{xy}^\mathrm{t}(H) = \rho_{xy}(H) - \eta(H)$, is plotted on the right-hand ordinate axis of Fig. \ref{fig:THE}(a).

The THE is negative, reaching an extremum at 400~mT of -31~n$\Omega$cm. The data show a strong resemblance to those of Huang and Chien,\cite{Huang2012} including the kink in $\rho_{xy}$ occurring from the THE, suggesting that the effect is repeatable, and robust in spite of the greater degree of crystallographic order in our film. The order of magnitude of the THE is similar to that reported by Huang and Chien for their 60~nm thick FeGe film, which was -46~n$\Omega$cm. Spin textures that give rise to a THE include skyrmion lattices in B20 crystals\cite{Lee2009,Neubauer2009,Kanazawa2011,Schulz2012,Huang2012,Li2013,
PorterArXiv2013} and fan-structures\cite{Shiomi2012}. Whatever spin texture is present in our film, it must be consistent with the MR $\propto -M^2$ scaling described in the previous section, but must also represent a departure from the simple conical state assumed in the model we present there, since that state has zero net topological winding number. Detailed studies of the spin textures in thin films of B20 materials are needed to resolve this point.

The THE extremum grows to $-136$~n$\Omega$cm as the temperature is raised to 150~K, which is perhaps indicative of a higher skyrmion winding number density than at 5 K, as shown in Fig. \ref{fig:THE}(b). Above this temperature the non-linearity in the ordinary Hall coefficient makes the fitting process unreliable. Overall, the THE remains negative over this temperature range, increasing in magnitude as the sample warms, as previously observed in an epitaxial sample\cite{Huang2012}. Thus, just as the resistivity $\rho_{xx}$ is quite insensitive to the fact that our sample is textured but not epitaxial (Fig. \ref{fig:rhoXXvT}), so is the topological Hall resistivity $\rho_{xy}^\mathrm{t}$. Thus, it is reasonable to expect that our findings about scattering mechanisms leading to MR (\S\ref{sec:Rxx}) and AHE (\S\ref{sec:AHE}) are likely also to apply to epitaxial layers.

\section{Summary}

The scattering mechanisms responsible for MR and the AHE have been determined for a thin film of (111) textured B20-ordered FeGe. Scattering from spin fluctuations is responsible for the large MR arising near to the helical ordering temperature, but as the temperature is reduced the contribution to scattering from electron-magnon processes becomes weaker. At these lowest temperatures, classical scattering and the Lorentz force from the magnetic field acting upon the carriers accounts for the MR above saturation and scattering from the conical state is at its strongest. The observed MR~$\propto -M^2$ scaling in the conical phase shows that a GMR-type model is appropriate to describe the spin-dependent scattering. The conical phase MR vanishes above $\sim 200$~K, well below the magnetic ordering temperature of $T_\mathrm{N} = 276$~K. Meanwhile $\rho_{xx}(H,T)$ is featureless at $T_\mathrm{N}$. The strong temperature dependence of the AHE scales as $\rho_{xx} \propto \rho_{xy}^2$, ruling out the skew scattering mechanism. Our results are consistent with the scaling theory of Onoda et al.,\cite{Onoda2008} which suggests that the AHE in our material is dominated by the intrinsic, Berry-phase, mechanism.

\begin{acknowledgments}
The authors would like to acknowledge financial support from the UK EPSRC (grant numbers EP/J007110/1, EP/K00512X/1, and EP/J021156/1).
\end{acknowledgments}


\begin{thebibliography}{62}%
\makeatletter
\providecommand \@ifxundefined [1]{%
 \@ifx{#1\undefined}
}%
\providecommand \@ifnum [1]{%
 \ifnum #1\expandafter \@firstoftwo
 \else \expandafter \@secondoftwo
 \fi
}%
\providecommand \@ifx [1]{%
 \ifx #1\expandafter \@firstoftwo
 \else \expandafter \@secondoftwo
 \fi
}%
\providecommand \natexlab [1]{#1}%
\providecommand \enquote  [1]{``#1''}%
\providecommand \bibnamefont  [1]{#1}%
\providecommand \bibfnamefont [1]{#1}%
\providecommand \citenamefont [1]{#1}%
\providecommand \href@noop [0]{\@secondoftwo}%
\providecommand \href [0]{\begingroup \@sanitize@url \@href}%
\providecommand \@href[1]{\@@startlink{#1}\@@href}%
\providecommand \@@href[1]{\endgroup#1\@@endlink}%
\providecommand \@sanitize@url [0]{\catcode `\\12\catcode `\$12\catcode
  `\&12\catcode `\#12\catcode `\^12\catcode `\_12\catcode `\%12\relax}%
\providecommand \@@startlink[1]{}%
\providecommand \@@endlink[0]{}%
\providecommand \url  [0]{\begingroup\@sanitize@url \@url }%
\providecommand \@url [1]{\endgroup\@href {#1}{\urlprefix }}%
\providecommand \urlprefix  [0]{URL }%
\providecommand \Eprint [0]{\href }%
\providecommand \doibase [0]{http://dx.doi.org/}%
\providecommand \selectlanguage [0]{\@gobble}%
\providecommand \bibinfo  [0]{\@secondoftwo}%
\providecommand \bibfield  [0]{\@secondoftwo}%
\providecommand \translation [1]{[#1]}%
\providecommand \BibitemOpen [0]{}%
\providecommand \bibitemStop [0]{}%
\providecommand \bibitemNoStop [0]{.\EOS\space}%
\providecommand \EOS [0]{\spacefactor3000\relax}%
\providecommand \BibitemShut  [1]{\csname bibitem#1\endcsname}%
\let\auto@bib@innerbib\@empty
\bibitem [{\citenamefont {Grigoriev}\ \emph {et~al.}(2006)\citenamefont
  {Grigoriev}, \citenamefont {Maleyev}, \citenamefont {Okorokov}, \citenamefont
  {Chetverikov}, \citenamefont {B\"oni}, \citenamefont {Georgii}, \citenamefont
  {Lamago}, \citenamefont {Eckerlebe},\ and\ \citenamefont
  {Pranzas}}]{Grigoriev2006}%
  \BibitemOpen
  \bibfield  {author} {\bibinfo {author} {\bibfnamefont {S.~V.}\ \bibnamefont
  {Grigoriev}}, \bibinfo {author} {\bibfnamefont {S.~V.}\ \bibnamefont
  {Maleyev}}, \bibinfo {author} {\bibfnamefont {A.~I.}\ \bibnamefont
  {Okorokov}}, \bibinfo {author} {\bibfnamefont {Y.~O.}\ \bibnamefont
  {Chetverikov}}, \bibinfo {author} {\bibfnamefont {P.}~\bibnamefont {B\"oni}},
  \bibinfo {author} {\bibfnamefont {R.}~\bibnamefont {Georgii}}, \bibinfo
  {author} {\bibfnamefont {D.}~\bibnamefont {Lamago}}, \bibinfo {author}
  {\bibfnamefont {H.}~\bibnamefont {Eckerlebe}}, \ and\ \bibinfo {author}
  {\bibfnamefont {K.}~\bibnamefont {Pranzas}},\ }\href {\doibase
  10.1103/PhysRevB.74.214414} {\bibfield  {journal} {\bibinfo  {journal} {Phys.
  Rev. B}\ }\textbf {\bibinfo {volume} {74}},\ \bibinfo {pages} {214414}
  (\bibinfo {year} {2006})}\BibitemShut {NoStop}%
\bibitem [{\citenamefont {M\"{u}hlbauer}\ \emph {et~al.}(2009)\citenamefont
  {M\"{u}hlbauer}, \citenamefont {Binz}, \citenamefont {Jonietz}, \citenamefont
  {Pfleiderer}, \citenamefont {Rosch}, \citenamefont {Neubauer}, \citenamefont
  {Georgii},\ and\ \citenamefont {B\"{o}ni}}]{Muhlbauer2009}%
  \BibitemOpen
  \bibfield  {author} {\bibinfo {author} {\bibfnamefont {S.}~\bibnamefont
  {M\"{u}hlbauer}}, \bibinfo {author} {\bibfnamefont {B.}~\bibnamefont {Binz}},
  \bibinfo {author} {\bibfnamefont {F.}~\bibnamefont {Jonietz}}, \bibinfo
  {author} {\bibfnamefont {C.}~\bibnamefont {Pfleiderer}}, \bibinfo {author}
  {\bibfnamefont {A.}~\bibnamefont {Rosch}}, \bibinfo {author} {\bibfnamefont
  {A.}~\bibnamefont {Neubauer}}, \bibinfo {author} {\bibfnamefont
  {R.}~\bibnamefont {Georgii}}, \ and\ \bibinfo {author} {\bibfnamefont
  {P.}~\bibnamefont {B\"{o}ni}},\ }\href {\doibase 10.1126/science.1166767}
  {\bibfield  {journal} {\bibinfo  {journal} {Science}\ }\textbf {\bibinfo
  {volume} {323}},\ \bibinfo {pages} {915} (\bibinfo {year}
  {2009})}\BibitemShut {NoStop}%
\bibitem [{\citenamefont {Karhu}\ \emph {et~al.}(2010)\citenamefont {Karhu},
  \citenamefont {Kahwaji}, \citenamefont {Monchesky}, \citenamefont {Parsons},
  \citenamefont {Robertson},\ and\ \citenamefont {Maunders}}]{Karhu2010}%
  \BibitemOpen
  \bibfield  {author} {\bibinfo {author} {\bibfnamefont {E.}~\bibnamefont
  {Karhu}}, \bibinfo {author} {\bibfnamefont {S.}~\bibnamefont {Kahwaji}},
  \bibinfo {author} {\bibfnamefont {T.~L.}\ \bibnamefont {Monchesky}}, \bibinfo
  {author} {\bibfnamefont {C.}~\bibnamefont {Parsons}}, \bibinfo {author}
  {\bibfnamefont {M.~D.}\ \bibnamefont {Robertson}}, \ and\ \bibinfo {author}
  {\bibfnamefont {C.}~\bibnamefont {Maunders}},\ }\href {\doibase
  10.1103/PhysRevB.82.184417} {\bibfield  {journal} {\bibinfo  {journal} {Phys.
  Rev. B}\ }\textbf {\bibinfo {volume} {82}},\ \bibinfo {pages} {184417}
  (\bibinfo {year} {2010})}\BibitemShut {NoStop}%
\bibitem [{\citenamefont {Wilson}\ \emph {et~al.}(2012)\citenamefont {Wilson},
  \citenamefont {Karhu}, \citenamefont {Quigley}, \citenamefont {R\"o\ss{}ler},
  \citenamefont {Butenko}, \citenamefont {Bogdanov}, \citenamefont
  {Robertson},\ and\ \citenamefont {Monchesky}}]{Wilson2012}%
  \BibitemOpen
  \bibfield  {author} {\bibinfo {author} {\bibfnamefont {M.~N.}\ \bibnamefont
  {Wilson}}, \bibinfo {author} {\bibfnamefont {E.~A.}\ \bibnamefont {Karhu}},
  \bibinfo {author} {\bibfnamefont {A.~S.}\ \bibnamefont {Quigley}}, \bibinfo
  {author} {\bibfnamefont {U.~K.}\ \bibnamefont {R\"o\ss{}ler}}, \bibinfo
  {author} {\bibfnamefont {A.~B.}\ \bibnamefont {Butenko}}, \bibinfo {author}
  {\bibfnamefont {A.~N.}\ \bibnamefont {Bogdanov}}, \bibinfo {author}
  {\bibfnamefont {M.~D.}\ \bibnamefont {Robertson}}, \ and\ \bibinfo {author}
  {\bibfnamefont {T.~L.}\ \bibnamefont {Monchesky}},\ }\href {\doibase
  10.1103/PhysRevB.86.144420} {\bibfield  {journal} {\bibinfo  {journal} {Phys.
  Rev. B}\ }\textbf {\bibinfo {volume} {86}},\ \bibinfo {pages} {144420}
  (\bibinfo {year} {2012})}\BibitemShut {NoStop}%
\bibitem [{\citenamefont {Li}\ \emph {et~al.}(2013)\citenamefont {Li},
  \citenamefont {Kanazawa}, \citenamefont {Yu}, \citenamefont {Tsukazaki},
  \citenamefont {Kawasaki}, \citenamefont {Ichikawa}, \citenamefont {Jin},
  \citenamefont {Kagawa},\ and\ \citenamefont {Tokura}}]{Li2013}%
  \BibitemOpen
  \bibfield  {author} {\bibinfo {author} {\bibfnamefont {Y.}~\bibnamefont
  {Li}}, \bibinfo {author} {\bibfnamefont {N.}~\bibnamefont {Kanazawa}},
  \bibinfo {author} {\bibfnamefont {X.~Z.}\ \bibnamefont {Yu}}, \bibinfo
  {author} {\bibfnamefont {A.}~\bibnamefont {Tsukazaki}}, \bibinfo {author}
  {\bibfnamefont {M.}~\bibnamefont {Kawasaki}}, \bibinfo {author}
  {\bibfnamefont {M.}~\bibnamefont {Ichikawa}}, \bibinfo {author}
  {\bibfnamefont {X.~F.}\ \bibnamefont {Jin}}, \bibinfo {author} {\bibfnamefont
  {F.}~\bibnamefont {Kagawa}}, \ and\ \bibinfo {author} {\bibfnamefont
  {Y.}~\bibnamefont {Tokura}},\ }\href {\doibase
  10.1103/PhysRevLett.110.117202} {\bibfield  {journal} {\bibinfo  {journal}
  {Phys. Rev. Lett.}\ }\textbf {\bibinfo {volume} {110}},\ \bibinfo {pages}
  {117202} (\bibinfo {year} {2013})}\BibitemShut {NoStop}%
\bibitem [{\citenamefont {Yu}\ \emph {et~al.}(2010)\citenamefont {Yu},
  \citenamefont {Onose}, \citenamefont {Kanazawa}, \citenamefont {Park},
  \citenamefont {Han}, \citenamefont {Matsui}, \citenamefont {Nagaosa},\ and\
  \citenamefont {Tokura}}]{Yu2010}%
  \BibitemOpen
  \bibfield  {author} {\bibinfo {author} {\bibfnamefont {X.~Z.}\ \bibnamefont
  {Yu}}, \bibinfo {author} {\bibfnamefont {Y.}~\bibnamefont {Onose}}, \bibinfo
  {author} {\bibfnamefont {N.}~\bibnamefont {Kanazawa}}, \bibinfo {author}
  {\bibfnamefont {J.~H.}\ \bibnamefont {Park}}, \bibinfo {author}
  {\bibfnamefont {J.~H.}\ \bibnamefont {Han}}, \bibinfo {author} {\bibfnamefont
  {Y.}~\bibnamefont {Matsui}}, \bibinfo {author} {\bibfnamefont
  {N.}~\bibnamefont {Nagaosa}}, \ and\ \bibinfo {author} {\bibfnamefont
  {Y.}~\bibnamefont {Tokura}},\ }\href@noop {} {\bibfield  {journal} {\bibinfo
  {journal} {Nature}\ }\textbf {\bibinfo {volume} {465}},\ \bibinfo {pages}
  {901} (\bibinfo {year} {2010})}\BibitemShut {NoStop}%
\bibitem [{\citenamefont {Grigoriev}\ \emph {et~al.}(2007)\citenamefont
  {Grigoriev}, \citenamefont {Dyadkin}, \citenamefont {Menzel}, \citenamefont
  {Schoenes}, \citenamefont {Chetverikov}, \citenamefont {Okorokov},
  \citenamefont {Eckerlebe},\ and\ \citenamefont {Maleyev}}]{Grigoriev2007}%
  \BibitemOpen
  \bibfield  {author} {\bibinfo {author} {\bibfnamefont {S.~V.}\ \bibnamefont
  {Grigoriev}}, \bibinfo {author} {\bibfnamefont {V.~A.}\ \bibnamefont
  {Dyadkin}}, \bibinfo {author} {\bibfnamefont {D.}~\bibnamefont {Menzel}},
  \bibinfo {author} {\bibfnamefont {J.}~\bibnamefont {Schoenes}}, \bibinfo
  {author} {\bibfnamefont {Y.~O.}\ \bibnamefont {Chetverikov}}, \bibinfo
  {author} {\bibfnamefont {A.~I.}\ \bibnamefont {Okorokov}}, \bibinfo {author}
  {\bibfnamefont {H.}~\bibnamefont {Eckerlebe}}, \ and\ \bibinfo {author}
  {\bibfnamefont {S.~V.}\ \bibnamefont {Maleyev}},\ }\href {\doibase
  10.1103/PhysRevB.76.224424} {\bibfield  {journal} {\bibinfo  {journal} {Phys.
  Rev. B}\ }\textbf {\bibinfo {volume} {76}},\ \bibinfo {pages} {224424}
  (\bibinfo {year} {2007})}\BibitemShut {NoStop}%
\bibitem [{\citenamefont {Porter}\ \emph {et~al.}(2012)\citenamefont {Porter},
  \citenamefont {Creeth},\ and\ \citenamefont {Marrows}}]{Porter2012}%
  \BibitemOpen
  \bibfield  {author} {\bibinfo {author} {\bibfnamefont {N.~A.}\ \bibnamefont
  {Porter}}, \bibinfo {author} {\bibfnamefont {G.~L.}\ \bibnamefont {Creeth}},
  \ and\ \bibinfo {author} {\bibfnamefont {C.~H.}\ \bibnamefont {Marrows}},\
  }\href {\doibase 10.1103/PhysRevB.86.064423} {\bibfield  {journal} {\bibinfo
  {journal} {Phys. Rev. B}\ }\textbf {\bibinfo {volume} {86}},\ \bibinfo
  {pages} {064423} (\bibinfo {year} {2012})}\BibitemShut {NoStop}%
\bibitem [{\citenamefont {Yu}\ \emph {et~al.}(2011)\citenamefont {Yu},
  \citenamefont {Kanazawa}, \citenamefont {Onose}, \citenamefont {Kimoto},
  \citenamefont {Zhang}, \citenamefont {Ishiwata}, \citenamefont {Matsui},\
  and\ \citenamefont {Tokura}}]{Yu2011}%
  \BibitemOpen
  \bibfield  {author} {\bibinfo {author} {\bibfnamefont {X.~Z.}\ \bibnamefont
  {Yu}}, \bibinfo {author} {\bibfnamefont {N.}~\bibnamefont {Kanazawa}},
  \bibinfo {author} {\bibfnamefont {Y.}~\bibnamefont {Onose}}, \bibinfo
  {author} {\bibfnamefont {K.}~\bibnamefont {Kimoto}}, \bibinfo {author}
  {\bibfnamefont {W.~Z.}\ \bibnamefont {Zhang}}, \bibinfo {author}
  {\bibfnamefont {S.}~\bibnamefont {Ishiwata}}, \bibinfo {author}
  {\bibfnamefont {Y.}~\bibnamefont {Matsui}}, \ and\ \bibinfo {author}
  {\bibfnamefont {Y.}~\bibnamefont {Tokura}},\ }\href {\doibase
  10.1038/nmat2916} {\bibfield  {journal} {\bibinfo  {journal} {Nature Mater.}\
  }\textbf {\bibinfo {volume} {10}},\ \bibinfo {pages} {106} (\bibinfo {year}
  {2011})}\BibitemShut {NoStop}%
\bibitem [{\citenamefont {Wilhelm}\ \emph {et~al.}(2011)\citenamefont
  {Wilhelm}, \citenamefont {Baenitz}, \citenamefont {Schmidt}, \citenamefont
  {R\"o\ss{}ler}, \citenamefont {Leonov},\ and\ \citenamefont
  {Bogdanov}}]{Wilhelm2011}%
  \BibitemOpen
  \bibfield  {author} {\bibinfo {author} {\bibfnamefont {H.}~\bibnamefont
  {Wilhelm}}, \bibinfo {author} {\bibfnamefont {M.}~\bibnamefont {Baenitz}},
  \bibinfo {author} {\bibfnamefont {M.}~\bibnamefont {Schmidt}}, \bibinfo
  {author} {\bibfnamefont {U.~K.}\ \bibnamefont {R\"o\ss{}ler}}, \bibinfo
  {author} {\bibfnamefont {A.~A.}\ \bibnamefont {Leonov}}, \ and\ \bibinfo
  {author} {\bibfnamefont {A.~N.}\ \bibnamefont {Bogdanov}},\ }\href {\doibase
  10.1103/PhysRevLett.107.127203} {\bibfield  {journal} {\bibinfo  {journal}
  {Phys. Rev. Lett.}\ }\textbf {\bibinfo {volume} {107}},\ \bibinfo {pages}
  {127203} (\bibinfo {year} {2011})}\BibitemShut {NoStop}%
\bibitem [{\citenamefont {Huang}\ and\ \citenamefont
  {Chien}(2012)}]{Huang2012}%
  \BibitemOpen
  \bibfield  {author} {\bibinfo {author} {\bibfnamefont {S.~X.}\ \bibnamefont
  {Huang}}\ and\ \bibinfo {author} {\bibfnamefont {C.~L.}\ \bibnamefont
  {Chien}},\ }\href {\doibase 10.1103/PhysRevLett.108.267201} {\bibfield
  {journal} {\bibinfo  {journal} {Phys. Rev. Lett.}\ }\textbf {\bibinfo
  {volume} {108}},\ \bibinfo {pages} {267201} (\bibinfo {year}
  {2012})}\BibitemShut {NoStop}%
\bibitem [{\citenamefont {Moskvin}\ \emph {et~al.}(2013)\citenamefont
  {Moskvin}, \citenamefont {Grigoriev}, \citenamefont {Dyadkin}, \citenamefont
  {Eckerlebe}, \citenamefont {Baenitz}, \citenamefont {Schmidt},\ and\
  \citenamefont {Wilhelm}}]{Moskvin2013}%
  \BibitemOpen
  \bibfield  {author} {\bibinfo {author} {\bibfnamefont {E.}~\bibnamefont
  {Moskvin}}, \bibinfo {author} {\bibfnamefont {S.}~\bibnamefont {Grigoriev}},
  \bibinfo {author} {\bibfnamefont {V.}~\bibnamefont {Dyadkin}}, \bibinfo
  {author} {\bibfnamefont {H.}~\bibnamefont {Eckerlebe}}, \bibinfo {author}
  {\bibfnamefont {M.}~\bibnamefont {Baenitz}}, \bibinfo {author} {\bibfnamefont
  {M.}~\bibnamefont {Schmidt}}, \ and\ \bibinfo {author} {\bibfnamefont
  {H.}~\bibnamefont {Wilhelm}},\ }\href {\doibase
  10.1103/PhysRevLett.110.077207} {\bibfield  {journal} {\bibinfo  {journal}
  {Phys. Rev. Lett.}\ }\textbf {\bibinfo {volume} {110}},\ \bibinfo {pages}
  {077207} (\bibinfo {year} {2013})}\BibitemShut {NoStop}%
\bibitem [{\citenamefont {Shibata}\ \emph {et~al.}(2013)\citenamefont
  {Shibata}, \citenamefont {Yu}, \citenamefont {Hara}, \citenamefont
  {Morikawa}, \citenamefont {Kanazawa}, \citenamefont {Kimoto}, \citenamefont
  {Ishiwata}, \citenamefont {Matsui},\ and\ \citenamefont
  {Tokura}}]{Shibata2013}%
  \BibitemOpen
  \bibfield  {author} {\bibinfo {author} {\bibfnamefont {K.}~\bibnamefont
  {Shibata}}, \bibinfo {author} {\bibfnamefont {X.~Z.}\ \bibnamefont {Yu}},
  \bibinfo {author} {\bibfnamefont {T.}~\bibnamefont {Hara}}, \bibinfo {author}
  {\bibfnamefont {D.}~\bibnamefont {Morikawa}}, \bibinfo {author}
  {\bibfnamefont {N.}~\bibnamefont {Kanazawa}}, \bibinfo {author}
  {\bibfnamefont {K.}~\bibnamefont {Kimoto}}, \bibinfo {author} {\bibfnamefont
  {S.}~\bibnamefont {Ishiwata}}, \bibinfo {author} {\bibfnamefont
  {Y.}~\bibnamefont {Matsui}}, \ and\ \bibinfo {author} {\bibfnamefont
  {Y.}~\bibnamefont {Tokura}},\ }\href@noop {} {\bibfield  {journal} {\bibinfo
  {journal} {Nature Nanotechnology}\ } (\bibinfo {year} {2013})}\BibitemShut
  {NoStop}%
\bibitem [{\citenamefont {Kanazawa}\ \emph {et~al.}(2011)\citenamefont
  {Kanazawa}, \citenamefont {Onose}, \citenamefont {Arima}, \citenamefont
  {Okuyama}, \citenamefont {Ohoyama}, \citenamefont {Wakimoto}, \citenamefont
  {Kakurai}, \citenamefont {Ishiwata},\ and\ \citenamefont
  {Tokura}}]{Kanazawa2011}%
  \BibitemOpen
  \bibfield  {author} {\bibinfo {author} {\bibfnamefont {N.}~\bibnamefont
  {Kanazawa}}, \bibinfo {author} {\bibfnamefont {Y.}~\bibnamefont {Onose}},
  \bibinfo {author} {\bibfnamefont {T.}~\bibnamefont {Arima}}, \bibinfo
  {author} {\bibfnamefont {D.}~\bibnamefont {Okuyama}}, \bibinfo {author}
  {\bibfnamefont {K.}~\bibnamefont {Ohoyama}}, \bibinfo {author} {\bibfnamefont
  {S.}~\bibnamefont {Wakimoto}}, \bibinfo {author} {\bibfnamefont
  {K.}~\bibnamefont {Kakurai}}, \bibinfo {author} {\bibfnamefont
  {S.}~\bibnamefont {Ishiwata}}, \ and\ \bibinfo {author} {\bibfnamefont
  {Y.}~\bibnamefont {Tokura}},\ }\href {\doibase
  10.1103/PhysRevLett.106.156603} {\bibfield  {journal} {\bibinfo  {journal}
  {Phys. Rev. Lett.}\ }\textbf {\bibinfo {volume} {106}},\ \bibinfo {pages}
  {156603} (\bibinfo {year} {2011})}\BibitemShut {NoStop}%
\bibitem [{\citenamefont {Kanazawa}\ \emph {et~al.}(2012)\citenamefont
  {Kanazawa}, \citenamefont {Kim}, \citenamefont {Inosov}, \citenamefont
  {White}, \citenamefont {Egetenmeyer}, \citenamefont {Gavilano}, \citenamefont
  {Ishiwata}, \citenamefont {Onose}, \citenamefont {Arima}, \citenamefont
  {Keimer},\ and\ \citenamefont {Tokura}}]{Kanazawa2012}%
  \BibitemOpen
  \bibfield  {author} {\bibinfo {author} {\bibfnamefont {N.}~\bibnamefont
  {Kanazawa}}, \bibinfo {author} {\bibfnamefont {J.-H.}\ \bibnamefont {Kim}},
  \bibinfo {author} {\bibfnamefont {D.~S.}\ \bibnamefont {Inosov}}, \bibinfo
  {author} {\bibfnamefont {J.~S.}\ \bibnamefont {White}}, \bibinfo {author}
  {\bibfnamefont {N.}~\bibnamefont {Egetenmeyer}}, \bibinfo {author}
  {\bibfnamefont {J.~L.}\ \bibnamefont {Gavilano}}, \bibinfo {author}
  {\bibfnamefont {S.}~\bibnamefont {Ishiwata}}, \bibinfo {author}
  {\bibfnamefont {Y.}~\bibnamefont {Onose}}, \bibinfo {author} {\bibfnamefont
  {T.}~\bibnamefont {Arima}}, \bibinfo {author} {\bibfnamefont
  {B.}~\bibnamefont {Keimer}}, \ and\ \bibinfo {author} {\bibfnamefont
  {Y.}~\bibnamefont {Tokura}},\ }\href {\doibase 10.1103/PhysRevB.86.134425}
  {\bibfield  {journal} {\bibinfo  {journal} {Phys. Rev. B}\ }\textbf {\bibinfo
  {volume} {86}},\ \bibinfo {pages} {134425} (\bibinfo {year}
  {2012})}\BibitemShut {NoStop}%
\bibitem [{\citenamefont {Ericsson}\ \emph {et~al.}(1981)\citenamefont
  {Ericsson}, \citenamefont {Karner}, \citenamefont {H\"{a}ggstr\"{o}m},\ and\
  \citenamefont {Chandra}}]{Ericsson1981}%
  \BibitemOpen
  \bibfield  {author} {\bibinfo {author} {\bibfnamefont {T.}~\bibnamefont
  {Ericsson}}, \bibinfo {author} {\bibfnamefont {W.}~\bibnamefont {Karner}},
  \bibinfo {author} {\bibfnamefont {L.}~\bibnamefont {H\"{a}ggstr\"{o}m}}, \
  and\ \bibinfo {author} {\bibfnamefont {K.}~\bibnamefont {Chandra}},\ }\href
  {http://stacks.iop.org/1402-4896/23/i=6/a=015} {\bibfield  {journal}
  {\bibinfo  {journal} {Physica Scripta}\ }\textbf {\bibinfo {volume} {23}},\
  \bibinfo {pages} {1118} (\bibinfo {year} {1981})}\BibitemShut {NoStop}%
\bibitem [{\citenamefont {Lebech}\ \emph {et~al.}(1989)\citenamefont {Lebech},
  \citenamefont {Bernhard},\ and\ \citenamefont {Freltoft}}]{Lebech1989}%
  \BibitemOpen
  \bibfield  {author} {\bibinfo {author} {\bibfnamefont {B.}~\bibnamefont
  {Lebech}}, \bibinfo {author} {\bibfnamefont {J.}~\bibnamefont {Bernhard}}, \
  and\ \bibinfo {author} {\bibfnamefont {T.}~\bibnamefont {Freltoft}},\ }\href
  {http://stacks.iop.org/0953-8984/1/i=35/a=010} {\bibfield  {journal}
  {\bibinfo  {journal} {J. Phys.-Condens. Mat.}\ }\textbf {\bibinfo {volume}
  {1}},\ \bibinfo {pages} {6105} (\bibinfo {year} {1989})}\BibitemShut
  {NoStop}%
\bibitem [{\citenamefont {Kiselev}\ \emph {et~al.}(2011)\citenamefont
  {Kiselev}, \citenamefont {Bogdanov}, \citenamefont {Sch\"{a}fer},\ and\
  \citenamefont {R\"{o}\ss{}ler}}]{Kiselev2011}%
  \BibitemOpen
  \bibfield  {author} {\bibinfo {author} {\bibfnamefont {N.~S.}\ \bibnamefont
  {Kiselev}}, \bibinfo {author} {\bibfnamefont {A.~N.}\ \bibnamefont
  {Bogdanov}}, \bibinfo {author} {\bibfnamefont {R.}~\bibnamefont
  {Sch\"{a}fer}}, \ and\ \bibinfo {author} {\bibfnamefont {U.~K.}\ \bibnamefont
  {R\"{o}\ss{}ler}},\ }\href {\doibase 10.1088/0022-3727/44/39/392001}
  {\bibfield  {journal} {\bibinfo  {journal} {J. Phys. D: Appl. Phys.}\
  }\textbf {\bibinfo {volume} {44}},\ \bibinfo {pages} {392001} (\bibinfo
  {year} {2011})}\BibitemShut {NoStop}%
\bibitem [{\citenamefont {Schulz}\ \emph {et~al.}(2012)\citenamefont {Schulz},
  \citenamefont {Ritz}, \citenamefont {Bauer}, \citenamefont {Halder},
  \citenamefont {Wagner}, \citenamefont {Franz}, \citenamefont {Pfleiderer},
  \citenamefont {Everschor}, \citenamefont {Garst},\ and\ \citenamefont
  {Rosch}}]{Schulz2012}%
  \BibitemOpen
  \bibfield  {author} {\bibinfo {author} {\bibfnamefont {T.}~\bibnamefont
  {Schulz}}, \bibinfo {author} {\bibfnamefont {R.}~\bibnamefont {Ritz}},
  \bibinfo {author} {\bibfnamefont {A.}~\bibnamefont {Bauer}}, \bibinfo
  {author} {\bibfnamefont {M.}~\bibnamefont {Halder}}, \bibinfo {author}
  {\bibfnamefont {M.}~\bibnamefont {Wagner}}, \bibinfo {author} {\bibfnamefont
  {C.}~\bibnamefont {Franz}}, \bibinfo {author} {\bibfnamefont
  {C.}~\bibnamefont {Pfleiderer}}, \bibinfo {author} {\bibfnamefont
  {K.}~\bibnamefont {Everschor}}, \bibinfo {author} {\bibfnamefont
  {M.}~\bibnamefont {Garst}}, \ and\ \bibinfo {author} {\bibfnamefont
  {A.}~\bibnamefont {Rosch}},\ }\href@noop {} {\bibfield  {journal} {\bibinfo
  {journal} {Nature Phys.}\ }\textbf {\bibinfo {volume} {8}},\ \bibinfo {pages}
  {301} (\bibinfo {year} {2012})}\BibitemShut {NoStop}%
\bibitem [{\citenamefont {Iwasaki}\ \emph {et~al.}(2013)\citenamefont
  {Iwasaki}, \citenamefont {Mochizuki},\ and\ \citenamefont
  {Nagaosa}}]{IwasakiNANO2013}%
  \BibitemOpen
  \bibfield  {author} {\bibinfo {author} {\bibfnamefont {J.}~\bibnamefont
  {Iwasaki}}, \bibinfo {author} {\bibfnamefont {M.}~\bibnamefont {Mochizuki}},
  \ and\ \bibinfo {author} {\bibfnamefont {N.}~\bibnamefont {Nagaosa}},\
  }\href@noop {} {\bibfield  {journal} {\bibinfo  {journal} {Nature Nano.}\
  }\textbf {\bibinfo {volume} {8}},\ \bibinfo {pages} {742} (\bibinfo {year}
  {2013})}\BibitemShut {NoStop}%
\bibitem [{\citenamefont {Fert}\ \emph {et~al.}(2013)\citenamefont {Fert},
  \citenamefont {Cros},\ and\ \citenamefont {Sampaio}}]{Fert2013}%
  \BibitemOpen
  \bibfield  {author} {\bibinfo {author} {\bibfnamefont {A.}~\bibnamefont
  {Fert}}, \bibinfo {author} {\bibfnamefont {V.}~\bibnamefont {Cros}}, \ and\
  \bibinfo {author} {\bibfnamefont {J.}~\bibnamefont {Sampaio}},\ }\href@noop
  {} {\bibfield  {journal} {\bibinfo  {journal} {Nature Nano.}\ }\textbf
  {\bibinfo {volume} {8}},\ \bibinfo {pages} {152} (\bibinfo {year}
  {2013})}\BibitemShut {NoStop}%
\bibitem [{\citenamefont {Karhu}\ \emph {et~al.}(2011)\citenamefont {Karhu},
  \citenamefont {Kahwaji}, \citenamefont {Robertson}, \citenamefont
  {Fritzsche}, \citenamefont {Kirby}, \citenamefont {Majkrzak},\ and\
  \citenamefont {Monchesky}}]{Karhu2011}%
  \BibitemOpen
  \bibfield  {author} {\bibinfo {author} {\bibfnamefont {E.~A.}\ \bibnamefont
  {Karhu}}, \bibinfo {author} {\bibfnamefont {S.}~\bibnamefont {Kahwaji}},
  \bibinfo {author} {\bibfnamefont {M.~D.}\ \bibnamefont {Robertson}}, \bibinfo
  {author} {\bibfnamefont {H.}~\bibnamefont {Fritzsche}}, \bibinfo {author}
  {\bibfnamefont {B.~J.}\ \bibnamefont {Kirby}}, \bibinfo {author}
  {\bibfnamefont {C.~F.}\ \bibnamefont {Majkrzak}}, \ and\ \bibinfo {author}
  {\bibfnamefont {T.~L.}\ \bibnamefont {Monchesky}},\ }\href {\doibase
  10.1103/PhysRevB.84.060404} {\bibfield  {journal} {\bibinfo  {journal} {Phys.
  Rev. B}\ }\textbf {\bibinfo {volume} {84}},\ \bibinfo {pages} {060404}
  (\bibinfo {year} {2011})}\BibitemShut {NoStop}%
\bibitem [{\citenamefont {Sinha}\ \emph {et~al.}(2014)\citenamefont {Sinha},
  \citenamefont {Porter},\ and\ \citenamefont {Marrows}}]{Sinha2014}%
  \BibitemOpen
  \bibfield  {author} {\bibinfo {author} {\bibfnamefont {P.}~\bibnamefont
  {Sinha}}, \bibinfo {author} {\bibfnamefont {N.~A.}\ \bibnamefont {Porter}}, \
  and\ \bibinfo {author} {\bibfnamefont {C.~H.}\ \bibnamefont {Marrows}},\
  }\href {\doibase 10.1103/PhysRevB.89.134426} {\bibfield  {journal} {\bibinfo
  {journal} {Phys. Rev. B}\ }\textbf {\bibinfo {volume} {89}},\ \bibinfo
  {pages} {134426} (\bibinfo {year} {2014})}\BibitemShut {NoStop}%
\bibitem [{\citenamefont {Morley}\ \emph {et~al.}(2011)\citenamefont {Morley},
  \citenamefont {Porter},\ and\ \citenamefont {Marrows}}]{Morley2011}%
  \BibitemOpen
  \bibfield  {author} {\bibinfo {author} {\bibfnamefont {S.~A.}\ \bibnamefont
  {Morley}}, \bibinfo {author} {\bibfnamefont {N.~A.}\ \bibnamefont {Porter}},
  \ and\ \bibinfo {author} {\bibfnamefont {C.~H.}\ \bibnamefont {Marrows}},\
  }\href {\doibase 10.1002/pssr.201105386} {\bibfield  {journal} {\bibinfo
  {journal} {Phys. Status Solidi-R}\ }\textbf {\bibinfo {volume} {5}},\
  \bibinfo {pages} {429} (\bibinfo {year} {2011})}\BibitemShut {NoStop}%
\bibitem [{\citenamefont {Nagaosa}\ \emph {et~al.}(2010)\citenamefont
  {Nagaosa}, \citenamefont {Sinova}, \citenamefont {Onoda}, \citenamefont
  {MacDonald},\ and\ \citenamefont {Ong}}]{Nagaosa2010}%
  \BibitemOpen
  \bibfield  {author} {\bibinfo {author} {\bibfnamefont {N.}~\bibnamefont
  {Nagaosa}}, \bibinfo {author} {\bibfnamefont {J.}~\bibnamefont {Sinova}},
  \bibinfo {author} {\bibfnamefont {S.}~\bibnamefont {Onoda}}, \bibinfo
  {author} {\bibfnamefont {A.~H.}\ \bibnamefont {MacDonald}}, \ and\ \bibinfo
  {author} {\bibfnamefont {N.~P.}\ \bibnamefont {Ong}},\ }\href {\doibase
  10.1103/RevModPhys.82.1539} {\bibfield  {journal} {\bibinfo  {journal} {Rev.
  Mod. Phys.}\ }\textbf {\bibinfo {volume} {82}},\ \bibinfo {pages} {1539}
  (\bibinfo {year} {2010})}\BibitemShut {NoStop}%
\bibitem [{\citenamefont {Manyala}\ \emph {et~al.}(2004)\citenamefont
  {Manyala}, \citenamefont {Sidis}, \citenamefont {Ditusa}, \citenamefont
  {Aeppli}, \citenamefont {Young},\ and\ \citenamefont {Fisk}}]{Manyala2004}%
  \BibitemOpen
  \bibfield  {author} {\bibinfo {author} {\bibfnamefont {N.}~\bibnamefont
  {Manyala}}, \bibinfo {author} {\bibfnamefont {Y.}~\bibnamefont {Sidis}},
  \bibinfo {author} {\bibfnamefont {J.~F.}\ \bibnamefont {Ditusa}}, \bibinfo
  {author} {\bibfnamefont {G.}~\bibnamefont {Aeppli}}, \bibinfo {author}
  {\bibfnamefont {D.~P.}\ \bibnamefont {Young}}, \ and\ \bibinfo {author}
  {\bibfnamefont {Z.}~\bibnamefont {Fisk}},\ }\href@noop {} {\bibfield
  {journal} {\bibinfo  {journal} {Nature Mater.}\ }\textbf {\bibinfo {volume}
  {3}},\ \bibinfo {pages} {255} (\bibinfo {year} {2004})}\BibitemShut {NoStop}%
\bibitem [{\citenamefont {Tian}\ \emph {et~al.}(2009)\citenamefont {Tian},
  \citenamefont {Ye},\ and\ \citenamefont {Jin}}]{Tian2009}%
  \BibitemOpen
  \bibfield  {author} {\bibinfo {author} {\bibfnamefont {Y.}~\bibnamefont
  {Tian}}, \bibinfo {author} {\bibfnamefont {L.}~\bibnamefont {Ye}}, \ and\
  \bibinfo {author} {\bibfnamefont {X.}~\bibnamefont {Jin}},\ }\href {\doibase
  10.1103/PhysRevLett.103.087206} {\bibfield  {journal} {\bibinfo  {journal}
  {Phys. Rev. Lett.}\ }\textbf {\bibinfo {volume} {103}},\ \bibinfo {pages}
  {087206} (\bibinfo {year} {2009})}\BibitemShut {NoStop}%
\bibitem [{\citenamefont {Ye}\ \emph {et~al.}(2012)\citenamefont {Ye},
  \citenamefont {Tian}, \citenamefont {Jin},\ and\ \citenamefont
  {Xiao}}]{Ye2012}%
  \BibitemOpen
  \bibfield  {author} {\bibinfo {author} {\bibfnamefont {L.}~\bibnamefont
  {Ye}}, \bibinfo {author} {\bibfnamefont {Y.}~\bibnamefont {Tian}}, \bibinfo
  {author} {\bibfnamefont {X.}~\bibnamefont {Jin}}, \ and\ \bibinfo {author}
  {\bibfnamefont {D.}~\bibnamefont {Xiao}},\ }\href {\doibase
  10.1103/PhysRevB.85.220403} {\bibfield  {journal} {\bibinfo  {journal} {Phys.
  Rev. B}\ }\textbf {\bibinfo {volume} {85}},\ \bibinfo {pages} {220403}
  (\bibinfo {year} {2012})}\BibitemShut {NoStop}%
\bibitem [{\citenamefont {Karplus}\ and\ \citenamefont
  {Luttinger}(1954)}]{Karplus1954}%
  \BibitemOpen
  \bibfield  {author} {\bibinfo {author} {\bibfnamefont {R.}~\bibnamefont
  {Karplus}}\ and\ \bibinfo {author} {\bibfnamefont {J.~M.}\ \bibnamefont
  {Luttinger}},\ }\href {\doibase 10.1103/PhysRev.95.1154} {\bibfield
  {journal} {\bibinfo  {journal} {Phys. Rev.}\ }\textbf {\bibinfo {volume}
  {95}},\ \bibinfo {pages} {1154} (\bibinfo {year} {1954})}\BibitemShut
  {NoStop}%
\bibitem [{\citenamefont {Chevrier}\ \emph {et~al.}(1992)\citenamefont
  {Chevrier} \emph {et~al.}}]{Chevrier1992}%
  \BibitemOpen
  \bibfield  {author} {\bibinfo {author} {\bibfnamefont {J.}~\bibnamefont
  {Chevrier}} \emph {et~al.},\ }\href {\doibase 10.1016/0169-4332(92)90267-2}
  {\bibfield  {journal} {\bibinfo  {journal} {Appl. Surf. Sci.}\ }\textbf
  {\bibinfo {volume} {56-58, Part 1}},\ \bibinfo {pages} {438 } (\bibinfo
  {year} {1992})}\BibitemShut {NoStop}%
\bibitem [{\citenamefont {Pedrazzini}\ \emph {et~al.}(2007)\citenamefont
  {Pedrazzini}, \citenamefont {Wilhelm}, \citenamefont {Jaccard}, \citenamefont
  {Jarlborg}, \citenamefont {Schmidt}, \citenamefont {Hanfland}, \citenamefont
  {Akselrud}, \citenamefont {Yuan}, \citenamefont {Schwarz}, \citenamefont
  {Grin},\ and\ \citenamefont {Steglich}}]{Pedrazzini2007}%
  \BibitemOpen
  \bibfield  {author} {\bibinfo {author} {\bibfnamefont {P.}~\bibnamefont
  {Pedrazzini}}, \bibinfo {author} {\bibfnamefont {H.}~\bibnamefont {Wilhelm}},
  \bibinfo {author} {\bibfnamefont {D.}~\bibnamefont {Jaccard}}, \bibinfo
  {author} {\bibfnamefont {T.}~\bibnamefont {Jarlborg}}, \bibinfo {author}
  {\bibfnamefont {M.}~\bibnamefont {Schmidt}}, \bibinfo {author} {\bibfnamefont
  {M.}~\bibnamefont {Hanfland}}, \bibinfo {author} {\bibfnamefont
  {L.}~\bibnamefont {Akselrud}}, \bibinfo {author} {\bibfnamefont {H.~Q.}\
  \bibnamefont {Yuan}}, \bibinfo {author} {\bibfnamefont {U.}~\bibnamefont
  {Schwarz}}, \bibinfo {author} {\bibfnamefont {Y.}~\bibnamefont {Grin}}, \
  and\ \bibinfo {author} {\bibfnamefont {F.}~\bibnamefont {Steglich}},\ }\href
  {\doibase 10.1103/PhysRevLett.98.047204} {\bibfield  {journal} {\bibinfo
  {journal} {Phys. Rev. Lett.}\ }\textbf {\bibinfo {volume} {98}},\ \bibinfo
  {pages} {047204} (\bibinfo {year} {2007})}\BibitemShut {NoStop}%
\bibitem [{\citenamefont {Yamada}\ \emph {et~al.}(2003)\citenamefont {Yamada},
  \citenamefont {Terao}, \citenamefont {Ohta},\ and\ \citenamefont
  {Kulatov}}]{Yamada2003}%
  \BibitemOpen
  \bibfield  {author} {\bibinfo {author} {\bibfnamefont {H.}~\bibnamefont
  {Yamada}}, \bibinfo {author} {\bibfnamefont {K.}~\bibnamefont {Terao}},
  \bibinfo {author} {\bibfnamefont {H.}~\bibnamefont {Ohta}}, \ and\ \bibinfo
  {author} {\bibfnamefont {E.}~\bibnamefont {Kulatov}},\ }\href {\doibase
  10.1016/S0921-4526(02)02471-7} {\bibfield  {journal} {\bibinfo  {journal}
  {Physica B: Condensed Matter}\ }\textbf {\bibinfo {volume} {329}},\ \bibinfo
  {pages} {1131 } (\bibinfo {year} {2003})}\BibitemShut {NoStop}%
\bibitem [{\citenamefont {Karhu}\ \emph {et~al.}(2012)\citenamefont {Karhu},
  \citenamefont {R\"o\ss{}ler}, \citenamefont {Bogdanov}, \citenamefont
  {Kahwaji}, \citenamefont {Kirby}, \citenamefont {Fritzsche}, \citenamefont
  {Robertson}, \citenamefont {Majkrzak},\ and\ \citenamefont
  {Monchesky}}]{Karhu2012}%
  \BibitemOpen
  \bibfield  {author} {\bibinfo {author} {\bibfnamefont {E.~A.}\ \bibnamefont
  {Karhu}}, \bibinfo {author} {\bibfnamefont {U.~K.}\ \bibnamefont
  {R\"o\ss{}ler}}, \bibinfo {author} {\bibfnamefont {A.~N.}\ \bibnamefont
  {Bogdanov}}, \bibinfo {author} {\bibfnamefont {S.}~\bibnamefont {Kahwaji}},
  \bibinfo {author} {\bibfnamefont {B.~J.}\ \bibnamefont {Kirby}}, \bibinfo
  {author} {\bibfnamefont {H.}~\bibnamefont {Fritzsche}}, \bibinfo {author}
  {\bibfnamefont {M.~D.}\ \bibnamefont {Robertson}}, \bibinfo {author}
  {\bibfnamefont {C.~F.}\ \bibnamefont {Majkrzak}}, \ and\ \bibinfo {author}
  {\bibfnamefont {T.~L.}\ \bibnamefont {Monchesky}},\ }\href {\doibase
  10.1103/PhysRevB.85.094429} {\bibfield  {journal} {\bibinfo  {journal} {Phys.
  Rev. B}\ }\textbf {\bibinfo {volume} {85}},\ \bibinfo {pages} {094429}
  (\bibinfo {year} {2012})}\BibitemShut {NoStop}%
\bibitem [{\citenamefont {Wilhelm}\ \emph {et~al.}(2012)\citenamefont
  {Wilhelm}, \citenamefont {Baenitz}, \citenamefont {Schmidt}, \citenamefont
  {Naylor}, \citenamefont {Lortz}, \citenamefont {R\"{o}\ss{}ler},
  \citenamefont {Leonov},\ and\ \citenamefont {Bogdanov}}]{Wilhelm2012}%
  \BibitemOpen
  \bibfield  {author} {\bibinfo {author} {\bibfnamefont {H.}~\bibnamefont
  {Wilhelm}}, \bibinfo {author} {\bibfnamefont {M.}~\bibnamefont {Baenitz}},
  \bibinfo {author} {\bibfnamefont {M.}~\bibnamefont {Schmidt}}, \bibinfo
  {author} {\bibfnamefont {C.}~\bibnamefont {Naylor}}, \bibinfo {author}
  {\bibfnamefont {R.}~\bibnamefont {Lortz}}, \bibinfo {author} {\bibfnamefont
  {U.~K.}\ \bibnamefont {R\"{o}\ss{}ler}}, \bibinfo {author} {\bibfnamefont
  {A.~A.}\ \bibnamefont {Leonov}}, \ and\ \bibinfo {author} {\bibfnamefont
  {A.~N.}\ \bibnamefont {Bogdanov}},\ }\href@noop {} {\bibfield  {journal}
  {\bibinfo  {journal} {J. Phys.: Condens. Matter}\ }\textbf {\bibinfo {volume}
  {24}},\ \bibinfo {pages} {294204} (\bibinfo {year} {2012})}\BibitemShut
  {NoStop}%
\bibitem [{\citenamefont {Uchida}\ \emph {et~al.}(2008)\citenamefont {Uchida},
  \citenamefont {Nagaosa}, \citenamefont {He}, \citenamefont {Kaneko},
  \citenamefont {Iguchi}, \citenamefont {Matsui},\ and\ \citenamefont
  {Tokura}}]{Uchida2008}%
  \BibitemOpen
  \bibfield  {author} {\bibinfo {author} {\bibfnamefont {M.}~\bibnamefont
  {Uchida}}, \bibinfo {author} {\bibfnamefont {N.}~\bibnamefont {Nagaosa}},
  \bibinfo {author} {\bibfnamefont {J.~P.}\ \bibnamefont {He}}, \bibinfo
  {author} {\bibfnamefont {Y.}~\bibnamefont {Kaneko}}, \bibinfo {author}
  {\bibfnamefont {S.}~\bibnamefont {Iguchi}}, \bibinfo {author} {\bibfnamefont
  {Y.}~\bibnamefont {Matsui}}, \ and\ \bibinfo {author} {\bibfnamefont
  {Y.}~\bibnamefont {Tokura}},\ }\href {\doibase 10.1103/PhysRevB.77.184402}
  {\bibfield  {journal} {\bibinfo  {journal} {Phys. Rev. B}\ }\textbf {\bibinfo
  {volume} {77}},\ \bibinfo {pages} {184402} (\bibinfo {year}
  {2008})}\BibitemShut {NoStop}%
\bibitem [{\citenamefont {Wilson}\ \emph {et~al.}(2013)\citenamefont {Wilson},
  \citenamefont {Karhu}, \citenamefont {Lake}, \citenamefont {Quigley},
  \citenamefont {Meynell}, \citenamefont {Bogdanov}, \citenamefont {Fritzsche},
  \citenamefont {R\"o\ss{}ler},\ and\ \citenamefont {Monchesky}}]{Wilson2013}%
  \BibitemOpen
  \bibfield  {author} {\bibinfo {author} {\bibfnamefont {M.~N.}\ \bibnamefont
  {Wilson}}, \bibinfo {author} {\bibfnamefont {E.~A.}\ \bibnamefont {Karhu}},
  \bibinfo {author} {\bibfnamefont {D.~P.}\ \bibnamefont {Lake}}, \bibinfo
  {author} {\bibfnamefont {A.~S.}\ \bibnamefont {Quigley}}, \bibinfo {author}
  {\bibfnamefont {S.}~\bibnamefont {Meynell}}, \bibinfo {author} {\bibfnamefont
  {A.~N.}\ \bibnamefont {Bogdanov}}, \bibinfo {author} {\bibfnamefont
  {H.}~\bibnamefont {Fritzsche}}, \bibinfo {author} {\bibfnamefont {U.~K.}\
  \bibnamefont {R\"o\ss{}ler}}, \ and\ \bibinfo {author} {\bibfnamefont
  {T.~L.}\ \bibnamefont {Monchesky}},\ }\href {\doibase
  10.1103/PhysRevB.88.214420} {\bibfield  {journal} {\bibinfo  {journal} {Phys.
  Rev. B}\ }\textbf {\bibinfo {volume} {88}},\ \bibinfo {pages} {214420}
  (\bibinfo {year} {2013})}\BibitemShut {NoStop}%
\bibitem [{\citenamefont {Rhodes}\ and\ \citenamefont
  {Wohlfarth}(1963)}]{Rhodes1963}%
  \BibitemOpen
  \bibfield  {author} {\bibinfo {author} {\bibfnamefont {P.}~\bibnamefont
  {Rhodes}}\ and\ \bibinfo {author} {\bibfnamefont {E.~P.}\ \bibnamefont
  {Wohlfarth}},\ }\href {\doibase 10.1098/rspa.1963.0086} {\bibfield  {journal}
  {\bibinfo  {journal} {Proc. R. Soc. Lond. A}\ }\textbf {\bibinfo {volume}
  {273}},\ \bibinfo {pages} {247} (\bibinfo {year} {1963})}\BibitemShut
  {NoStop}%
\bibitem [{\citenamefont {Pippard}(1989)}]{PippardBook1989}%
  \BibitemOpen
  \bibfield  {author} {\bibinfo {author} {\bibfnamefont {A.~B.}\ \bibnamefont
  {Pippard}},\ }\href@noop {} {\emph {\bibinfo {title} {Magnetoresistance in
  Metals}}}\ (\bibinfo  {publisher} {Cambridge University Press},\ \bibinfo
  {address} {Cambridge},\ \bibinfo {year} {1989})\BibitemShut {NoStop}%
\bibitem [{\citenamefont {Yosida}(1957)}]{Yosida1957}%
  \BibitemOpen
  \bibfield  {author} {\bibinfo {author} {\bibfnamefont {K.}~\bibnamefont
  {Yosida}},\ }\href@noop {} {\bibfield  {journal} {\bibinfo  {journal} {Phys.
  Rev.}\ }\textbf {\bibinfo {volume} {107}},\ \bibinfo {pages} {396} (\bibinfo
  {year} {1957})}\BibitemShut {NoStop}%
\bibitem [{\citenamefont {Khosla}\ and\ \citenamefont
  {Fischer}(1970)}]{Khosla1970}%
  \BibitemOpen
  \bibfield  {author} {\bibinfo {author} {\bibfnamefont {R.~P.}\ \bibnamefont
  {Khosla}}\ and\ \bibinfo {author} {\bibfnamefont {J.~R.}\ \bibnamefont
  {Fischer}},\ }\href {\doibase 10.1103/PhysRevB.2.4084} {\bibfield  {journal}
  {\bibinfo  {journal} {Phys. Rev. B}\ }\textbf {\bibinfo {volume} {2}},\
  \bibinfo {pages} {4084} (\bibinfo {year} {1970})}\BibitemShut {NoStop}%
\bibitem [{\citenamefont {Raquet}\ \emph {et~al.}(2002)\citenamefont {Raquet},
  \citenamefont {Viret}, \citenamefont {Sondergard}, \citenamefont {Cespedes},\
  and\ \citenamefont {Mamy}}]{Raquet2002}%
  \BibitemOpen
  \bibfield  {author} {\bibinfo {author} {\bibfnamefont {B.}~\bibnamefont
  {Raquet}}, \bibinfo {author} {\bibfnamefont {M.}~\bibnamefont {Viret}},
  \bibinfo {author} {\bibfnamefont {E.}~\bibnamefont {Sondergard}}, \bibinfo
  {author} {\bibfnamefont {O.}~\bibnamefont {Cespedes}}, \ and\ \bibinfo
  {author} {\bibfnamefont {R.}~\bibnamefont {Mamy}},\ }\href {\doibase
  10.1103/PhysRevB.66.024433} {\bibfield  {journal} {\bibinfo  {journal} {Phys.
  Rev. B}\ }\textbf {\bibinfo {volume} {66}},\ \bibinfo {pages} {024433}
  (\bibinfo {year} {2002})}\BibitemShut {NoStop}%
\bibitem [{\citenamefont {Marrows}\ and\ \citenamefont
  {Dalton}(2004)}]{Marrows2004}%
  \BibitemOpen
  \bibfield  {author} {\bibinfo {author} {\bibfnamefont {C.~H.}\ \bibnamefont
  {Marrows}}\ and\ \bibinfo {author} {\bibfnamefont {B.~C.}\ \bibnamefont
  {Dalton}},\ }\href@noop {} {\bibfield  {journal} {\bibinfo  {journal} {Phys.
  Rev. Lett.}\ }\textbf {\bibinfo {volume} {92}},\ \bibinfo {pages} {097206}
  (\bibinfo {year} {2004})}\BibitemShut {NoStop}%
\bibitem [{\citenamefont {Ishikawa}\ \emph {et~al.}(1977)\citenamefont
  {Ishikawa}, \citenamefont {Shirane}, \citenamefont {Tarvin},\ and\
  \citenamefont {Kohgi}}]{Ishikawa1977}%
  \BibitemOpen
  \bibfield  {author} {\bibinfo {author} {\bibfnamefont {Y.}~\bibnamefont
  {Ishikawa}}, \bibinfo {author} {\bibfnamefont {G.}~\bibnamefont {Shirane}},
  \bibinfo {author} {\bibfnamefont {J.~A.}\ \bibnamefont {Tarvin}}, \ and\
  \bibinfo {author} {\bibfnamefont {M.}~\bibnamefont {Kohgi}},\ }\href
  {\doibase 10.1103/PhysRevB.16.4956} {\bibfield  {journal} {\bibinfo
  {journal} {Phys. Rev. B}\ }\textbf {\bibinfo {volume} {16}},\ \bibinfo
  {pages} {4956} (\bibinfo {year} {1977})}\BibitemShut {NoStop}%
\bibitem [{\citenamefont {Taylor}\ \emph {et~al.}(1968)\citenamefont {Taylor},
  \citenamefont {Isin},\ and\ \citenamefont {Coleman}}]{Taylor1968}%
  \BibitemOpen
  \bibfield  {author} {\bibinfo {author} {\bibfnamefont {G.~R.}\ \bibnamefont
  {Taylor}}, \bibinfo {author} {\bibfnamefont {A.}~\bibnamefont {Isin}}, \ and\
  \bibinfo {author} {\bibfnamefont {R.~V.}\ \bibnamefont {Coleman}},\ }\href
  {\doibase 10.1103/PhysRev.165.621} {\bibfield  {journal} {\bibinfo  {journal}
  {Phys. Rev.}\ }\textbf {\bibinfo {volume} {165}},\ \bibinfo {pages} {621}
  (\bibinfo {year} {1968})}\BibitemShut {NoStop}%
\bibitem [{\citenamefont {Janoschek}\ \emph {et~al.}(2013)\citenamefont
  {Janoschek}, \citenamefont {Garst}, \citenamefont {Bauer}, \citenamefont
  {Krautscheid}, \citenamefont {Georgii}, \citenamefont {B\"oni},\ and\
  \citenamefont {Pfleiderer}}]{Janoschek2013}%
  \BibitemOpen
  \bibfield  {author} {\bibinfo {author} {\bibfnamefont {M.}~\bibnamefont
  {Janoschek}}, \bibinfo {author} {\bibfnamefont {M.}~\bibnamefont {Garst}},
  \bibinfo {author} {\bibfnamefont {A.}~\bibnamefont {Bauer}}, \bibinfo
  {author} {\bibfnamefont {P.}~\bibnamefont {Krautscheid}}, \bibinfo {author}
  {\bibfnamefont {R.}~\bibnamefont {Georgii}}, \bibinfo {author} {\bibfnamefont
  {P.}~\bibnamefont {B\"oni}}, \ and\ \bibinfo {author} {\bibfnamefont
  {C.}~\bibnamefont {Pfleiderer}},\ }\href {\doibase
  10.1103/PhysRevB.87.134407} {\bibfield  {journal} {\bibinfo  {journal} {Phys.
  Rev. B}\ }\textbf {\bibinfo {volume} {87}},\ \bibinfo {pages} {134407}
  (\bibinfo {year} {2013})}\BibitemShut {NoStop}%
\bibitem [{\citenamefont {Kadowaki}\ \emph {et~al.}(1982)\citenamefont
  {Kadowaki}, \citenamefont {Okuda},\ and\ \citenamefont
  {Date}}]{Kadowaki1982}%
  \BibitemOpen
  \bibfield  {author} {\bibinfo {author} {\bibfnamefont {K.}~\bibnamefont
  {Kadowaki}}, \bibinfo {author} {\bibfnamefont {K.}~\bibnamefont {Okuda}}, \
  and\ \bibinfo {author} {\bibfnamefont {M.}~\bibnamefont {Date}},\ }\href
  {\doibase 10.1143/JPSJ.51.2433} {\bibfield  {journal} {\bibinfo  {journal}
  {J. Phys. Soc. Jpn.}\ }\textbf {\bibinfo {volume} {51}},\ \bibinfo {pages}
  {2433} (\bibinfo {year} {1982})}\BibitemShut {NoStop}%
\bibitem [{\citenamefont {Sakakibara}\ \emph {et~al.}(1982)\citenamefont
  {Sakakibara}, \citenamefont {Mollymoto},\ and\ \citenamefont
  {Date}}]{Sakakibara1982}%
  \BibitemOpen
  \bibfield  {author} {\bibinfo {author} {\bibfnamefont {T.}~\bibnamefont
  {Sakakibara}}, \bibinfo {author} {\bibfnamefont {H.}~\bibnamefont
  {Mollymoto}}, \ and\ \bibinfo {author} {\bibfnamefont {M.}~\bibnamefont
  {Date}},\ }\href {\doibase 10.1143/JPSJ.51.2439} {\bibfield  {journal}
  {\bibinfo  {journal} {J. Phys. Soc. Jpn.}\ }\textbf {\bibinfo {volume}
  {51}},\ \bibinfo {pages} {2439} (\bibinfo {year} {1982})}\BibitemShut
  {NoStop}%
\bibitem [{\citenamefont {Demishev}\ \emph {et~al.}(2012)\citenamefont
  {Demishev}, \citenamefont {Glushkov}, \citenamefont {Lobanova}, \citenamefont
  {Anisimov}, \citenamefont {Ivanov}, \citenamefont {Ishchenko}, \citenamefont
  {Karasev}, \citenamefont {Samarin}, \citenamefont {Sluchanko}, \citenamefont
  {Zimin},\ and\ \citenamefont {Semeno}}]{Demishev2012}%
  \BibitemOpen
  \bibfield  {author} {\bibinfo {author} {\bibfnamefont {S.~V.}\ \bibnamefont
  {Demishev}}, \bibinfo {author} {\bibfnamefont {V.~V.}\ \bibnamefont
  {Glushkov}}, \bibinfo {author} {\bibfnamefont {I.~I.}\ \bibnamefont
  {Lobanova}}, \bibinfo {author} {\bibfnamefont {M.~A.}\ \bibnamefont
  {Anisimov}}, \bibinfo {author} {\bibfnamefont {V.~Y.}\ \bibnamefont
  {Ivanov}}, \bibinfo {author} {\bibfnamefont {T.~V.}\ \bibnamefont
  {Ishchenko}}, \bibinfo {author} {\bibfnamefont {M.~S.}\ \bibnamefont
  {Karasev}}, \bibinfo {author} {\bibfnamefont {N.~A.}\ \bibnamefont
  {Samarin}}, \bibinfo {author} {\bibfnamefont {N.~E.}\ \bibnamefont
  {Sluchanko}}, \bibinfo {author} {\bibfnamefont {V.~M.}\ \bibnamefont
  {Zimin}}, \ and\ \bibinfo {author} {\bibfnamefont {A.~V.}\ \bibnamefont
  {Semeno}},\ }\href {\doibase 10.1103/PhysRevB.85.045131} {\bibfield
  {journal} {\bibinfo  {journal} {Phys. Rev. B}\ }\textbf {\bibinfo {volume}
  {85}},\ \bibinfo {pages} {045131} (\bibinfo {year} {2012})}\BibitemShut
  {NoStop}%
\bibitem [{\citenamefont {Marrows}(2005)}]{Marrows2005}%
  \BibitemOpen
  \bibfield  {author} {\bibinfo {author} {\bibfnamefont {C.~H.}\ \bibnamefont
  {Marrows}},\ }\href {\doibase 10.1080/00018730500442209} {\bibfield
  {journal} {\bibinfo  {journal} {Advances in Physics}\ }\textbf {\bibinfo
  {volume} {54}},\ \bibinfo {pages} {585} (\bibinfo {year} {2005})}\BibitemShut
  {NoStop}%
\bibitem [{\citenamefont {Levy}\ and\ \citenamefont {Zhang}(1997)}]{Levy1997}%
  \BibitemOpen
  \bibfield  {author} {\bibinfo {author} {\bibfnamefont {P.~M.}\ \bibnamefont
  {Levy}}\ and\ \bibinfo {author} {\bibfnamefont {S.}~\bibnamefont {Zhang}},\
  }\href@noop {} {\bibfield  {journal} {\bibinfo  {journal} {Phys. Rev. Lett.}\
  }\textbf {\bibinfo {volume} {79}},\ \bibinfo {pages} {5110} (\bibinfo {year}
  {1997})}\BibitemShut {NoStop}%
\bibitem [{\citenamefont {Nagaosa}\ and\ \citenamefont
  {Tokura}(2012)}]{Nagaosa2012}%
  \BibitemOpen
  \bibfield  {author} {\bibinfo {author} {\bibfnamefont {N.}~\bibnamefont
  {Nagaosa}}\ and\ \bibinfo {author} {\bibfnamefont {Y.}~\bibnamefont
  {Tokura}},\ }\href@noop {} {\bibfield  {journal} {\bibinfo  {journal}
  {Physica Scripta}\ }\textbf {\bibinfo {volume} {2012}},\ \bibinfo {pages}
  {014020} (\bibinfo {year} {2012})}\BibitemShut {NoStop}%
\bibitem [{\citenamefont {Bruno}\ \emph {et~al.}(2004)\citenamefont {Bruno},
  \citenamefont {Dugaev},\ and\ \citenamefont {Taillefumier}}]{Bruno2004}%
  \BibitemOpen
  \bibfield  {author} {\bibinfo {author} {\bibfnamefont {P.}~\bibnamefont
  {Bruno}}, \bibinfo {author} {\bibfnamefont {V.~K.}\ \bibnamefont {Dugaev}}, \
  and\ \bibinfo {author} {\bibfnamefont {M.}~\bibnamefont {Taillefumier}},\
  }\href {\doibase 10.1103/PhysRevLett.93.096806} {\bibfield  {journal}
  {\bibinfo  {journal} {Phys. Rev. Lett.}\ }\textbf {\bibinfo {volume} {93}},\
  \bibinfo {pages} {096806} (\bibinfo {year} {2004})}\BibitemShut {NoStop}%
\bibitem [{\citenamefont {Tatara}\ \emph {et~al.}(2008)\citenamefont {Tatara},
  \citenamefont {Kohno},\ and\ \citenamefont {Shibata}}]{Tatara2008}%
  \BibitemOpen
  \bibfield  {author} {\bibinfo {author} {\bibfnamefont {G.}~\bibnamefont
  {Tatara}}, \bibinfo {author} {\bibfnamefont {H.}~\bibnamefont {Kohno}}, \
  and\ \bibinfo {author} {\bibfnamefont {J.}~\bibnamefont {Shibata}},\ }\href
  {\doibase 10.1016/j.physrep.2008.07.003} {\bibfield  {journal} {\bibinfo
  {journal} {Physics Reports}\ }\textbf {\bibinfo {volume} {468}},\ \bibinfo
  {pages} {213} (\bibinfo {year} {2008})}\BibitemShut {NoStop}%
\bibitem [{\citenamefont {Hurd}(1972)}]{HurdHallBook1972}%
  \BibitemOpen
  \bibfield  {author} {\bibinfo {author} {\bibfnamefont {C.~M.}\ \bibnamefont
  {Hurd}},\ }\href@noop {} {\emph {\bibinfo {title} {The Hall Effect in Metals
  and Alloys}}}\ (\bibinfo  {publisher} {Plenum Press},\ \bibinfo {address}
  {New York},\ \bibinfo {year} {1972})\BibitemShut {NoStop}%
\bibitem [{\citenamefont {Kim}(1999)}]{Kim1999}%
  \BibitemOpen
  \bibfield  {author} {\bibinfo {author} {\bibfnamefont {J.~S.}\ \bibnamefont
  {Kim}},\ }\href {\doibase 10.1063/1.371187} {\bibfield  {journal} {\bibinfo
  {journal} {J. Appl. Phys.}\ }\textbf {\bibinfo {volume} {86}},\ \bibinfo
  {pages} {3187} (\bibinfo {year} {1999})}\BibitemShut {NoStop}%
\bibitem [{\citenamefont {Yokouchi}\ \emph {et~al.}(2014)\citenamefont
  {Yokouchi}, \citenamefont {Kanazawa}, \citenamefont {Tsukazaki},
  \citenamefont {Kozuka}, \citenamefont {Kawasaki}, \citenamefont {Ichikawa},
  \citenamefont {Kagawa},\ and\ \citenamefont {Tokura}}]{Yokouchi2014}%
  \BibitemOpen
  \bibfield  {author} {\bibinfo {author} {\bibfnamefont {T.}~\bibnamefont
  {Yokouchi}}, \bibinfo {author} {\bibfnamefont {N.}~\bibnamefont {Kanazawa}},
  \bibinfo {author} {\bibfnamefont {A.}~\bibnamefont {Tsukazaki}}, \bibinfo
  {author} {\bibfnamefont {Y.}~\bibnamefont {Kozuka}}, \bibinfo {author}
  {\bibfnamefont {M.}~\bibnamefont {Kawasaki}}, \bibinfo {author}
  {\bibfnamefont {M.}~\bibnamefont {Ichikawa}}, \bibinfo {author}
  {\bibfnamefont {F.}~\bibnamefont {Kagawa}}, \ and\ \bibinfo {author}
  {\bibfnamefont {Y.}~\bibnamefont {Tokura}},\ }\href {\doibase
  10.1103/PhysRevB.89.064416} {\bibfield  {journal} {\bibinfo  {journal} {Phys.
  Rev. B}\ }\textbf {\bibinfo {volume} {89}},\ \bibinfo {pages} {064416}
  (\bibinfo {year} {2014})}\BibitemShut {NoStop}%
\bibitem [{\citenamefont {Zeng}\ \emph {et~al.}(2006)\citenamefont {Zeng},
  \citenamefont {Yao}, \citenamefont {Niu},\ and\ \citenamefont
  {Weitering}}]{Zeng2006}%
  \BibitemOpen
  \bibfield  {author} {\bibinfo {author} {\bibfnamefont {C.}~\bibnamefont
  {Zeng}}, \bibinfo {author} {\bibfnamefont {Y.}~\bibnamefont {Yao}}, \bibinfo
  {author} {\bibfnamefont {Q.}~\bibnamefont {Niu}}, \ and\ \bibinfo {author}
  {\bibfnamefont {H.~H.}\ \bibnamefont {Weitering}},\ }\href {\doibase
  10.1103/PhysRevLett.96.037204} {\bibfield  {journal} {\bibinfo  {journal}
  {Phys. Rev. Lett.}\ }\textbf {\bibinfo {volume} {96}},\ \bibinfo {pages}
  {037204} (\bibinfo {year} {2006})}\BibitemShut {NoStop}%
\bibitem [{\citenamefont {Onoda}\ \emph {et~al.}(2008)\citenamefont {Onoda},
  \citenamefont {Sugimoto},\ and\ \citenamefont {Nagaosa}}]{Onoda2008}%
  \BibitemOpen
  \bibfield  {author} {\bibinfo {author} {\bibfnamefont {S.}~\bibnamefont
  {Onoda}}, \bibinfo {author} {\bibfnamefont {N.}~\bibnamefont {Sugimoto}}, \
  and\ \bibinfo {author} {\bibfnamefont {N.}~\bibnamefont {Nagaosa}},\
  }\href@noop {} {\bibfield  {journal} {\bibinfo  {journal} {Phys. Rev. B}\
  }\textbf {\bibinfo {volume} {77}},\ \bibinfo {pages} {165103} (\bibinfo
  {year} {2008})}\BibitemShut {NoStop}%
\bibitem [{\citenamefont {Lee}\ \emph {et~al.}(2009)\citenamefont {Lee},
  \citenamefont {Kang}, \citenamefont {Onose}, \citenamefont {Tokura},\ and\
  \citenamefont {Ong}}]{Lee2009}%
  \BibitemOpen
  \bibfield  {author} {\bibinfo {author} {\bibfnamefont {M.}~\bibnamefont
  {Lee}}, \bibinfo {author} {\bibfnamefont {W.}~\bibnamefont {Kang}}, \bibinfo
  {author} {\bibfnamefont {Y.}~\bibnamefont {Onose}}, \bibinfo {author}
  {\bibfnamefont {Y.}~\bibnamefont {Tokura}}, \ and\ \bibinfo {author}
  {\bibfnamefont {N.~P.}\ \bibnamefont {Ong}},\ }\href {\doibase
  10.1103/PhysRevLett.102.186601} {\bibfield  {journal} {\bibinfo  {journal}
  {Phys. Rev. Lett.}\ }\textbf {\bibinfo {volume} {102}},\ \bibinfo {pages}
  {186601} (\bibinfo {year} {2009})}\BibitemShut {NoStop}%
\bibitem [{\citenamefont {Neubauer}\ \emph {et~al.}(2009)\citenamefont
  {Neubauer}, \citenamefont {Pfleiderer}, \citenamefont {Binz}, \citenamefont
  {Rosch}, \citenamefont {Ritz}, \citenamefont {Niklowitz},\ and\ \citenamefont
  {B\"oni}}]{Neubauer2009}%
  \BibitemOpen
  \bibfield  {author} {\bibinfo {author} {\bibfnamefont {A.}~\bibnamefont
  {Neubauer}}, \bibinfo {author} {\bibfnamefont {C.}~\bibnamefont
  {Pfleiderer}}, \bibinfo {author} {\bibfnamefont {B.}~\bibnamefont {Binz}},
  \bibinfo {author} {\bibfnamefont {A.}~\bibnamefont {Rosch}}, \bibinfo
  {author} {\bibfnamefont {R.}~\bibnamefont {Ritz}}, \bibinfo {author}
  {\bibfnamefont {P.~G.}\ \bibnamefont {Niklowitz}}, \ and\ \bibinfo {author}
  {\bibfnamefont {P.}~\bibnamefont {B\"oni}},\ }\href {\doibase
  10.1103/PhysRevLett.102.186602} {\bibfield  {journal} {\bibinfo  {journal}
  {Phys. Rev. Lett.}\ }\textbf {\bibinfo {volume} {102}},\ \bibinfo {pages}
  {186602} (\bibinfo {year} {2009})}\BibitemShut {NoStop}%
\bibitem [{\citenamefont {{Porter}}\ \emph {et~al.}(2013)\citenamefont
  {{Porter}}, \citenamefont {{Sinha}}, \citenamefont {{Ward}}, \citenamefont
  {{Dobrynin}}, \citenamefont {{Brydson}}, \citenamefont {{Charlton}},
  \citenamefont {{Kinane}}, \citenamefont {{Robertson}}, \citenamefont
  {{Langridge}},\ and\ \citenamefont {{Marrows}}}]{PorterArXiv2013}%
  \BibitemOpen
  \bibfield  {author} {\bibinfo {author} {\bibfnamefont {N.~A.}\ \bibnamefont
  {{Porter}}}, \bibinfo {author} {\bibfnamefont {P.}~\bibnamefont {{Sinha}}},
  \bibinfo {author} {\bibfnamefont {M.~B.}\ \bibnamefont {{Ward}}}, \bibinfo
  {author} {\bibfnamefont {A.~N.}\ \bibnamefont {{Dobrynin}}}, \bibinfo
  {author} {\bibfnamefont {R.~M.~D.}\ \bibnamefont {{Brydson}}}, \bibinfo
  {author} {\bibfnamefont {T.~R.}\ \bibnamefont {{Charlton}}}, \bibinfo
  {author} {\bibfnamefont {C.~J.}\ \bibnamefont {{Kinane}}}, \bibinfo {author}
  {\bibfnamefont {M.~D.}\ \bibnamefont {{Robertson}}}, \bibinfo {author}
  {\bibfnamefont {S.}~\bibnamefont {{Langridge}}}, \ and\ \bibinfo {author}
  {\bibfnamefont {C.~H.}\ \bibnamefont {{Marrows}}},\ }\href@noop {} {\
  (\bibinfo {year} {2013})},\ \Eprint {http://arxiv.org/abs/1312.1722}
  {arXiv:1312.1722 [cond-mat.mes-hall]} \BibitemShut {NoStop}%
\bibitem [{\citenamefont {Shiomi}\ \emph {et~al.}(2012)\citenamefont {Shiomi},
  \citenamefont {Iguchi},\ and\ \citenamefont {Tokura}}]{Shiomi2012}%
  \BibitemOpen
  \bibfield  {author} {\bibinfo {author} {\bibfnamefont {Y.}~\bibnamefont
  {Shiomi}}, \bibinfo {author} {\bibfnamefont {S.}~\bibnamefont {Iguchi}}, \
  and\ \bibinfo {author} {\bibfnamefont {Y.}~\bibnamefont {Tokura}},\ }\href
  {\doibase 10.1103/PhysRevB.86.180404} {\bibfield  {journal} {\bibinfo
  {journal} {Phys. Rev. B}\ }\textbf {\bibinfo {volume} {86}},\ \bibinfo
  {pages} {180404} (\bibinfo {year} {2012})}\BibitemShut {NoStop}%
\end{thebibliography}
%

\end{document}